\documentclass[aps,pra,twocolumn,showpacs,superscriptaddress]{revtex4}
\usepackage[T2A]{fontenc}
\usepackage[cp1251]{inputenc}
\usepackage[english]{babel}
\usepackage{graphicx}
\usepackage{color}
\usepackage{bm}
\usepackage{amssymb}
\usepackage[intlimits]{amsmath}
\usepackage{amsmath}

\newcommand{\be}{\begin{equation}}
\newcommand{\ee}{\end{equation}}
\newcommand{\ben}{\begin{equation*}}
\newcommand{\een}{\end{equation*}}

\begin{document}

\hskip 2cm
\date{\today}

\title{Calculation of mutual information for nonlinear communication channel at large SNR}

\author{I.~S.~Terekhov}
\email[E-mail: ]{I.S.Terekhov@gmail.com} \affiliation{Budker Institute of  Nuclear
Physics of Siberian Branch Russian  Academy of Sciences, Novosibirsk, 630090 Russia}
 \affiliation{Novosibirsk State University, Novosibirsk, 630090 Russia}
\author{A.~V.~Reznichenko}
\email[E-mail: ]{A.V.Reznichenko@inp.nsk.su} \affiliation{Budker Institute of
Nuclear Physics of Siberian Branch Russian Academy of Sciences, Novosibirsk, 630090
Russia} \affiliation{Novosibirsk State University, Novosibirsk, 630090 Russia}
\author{S.~K.~Turitsyn}
\email[E-mail: ]{s.k.turitsyn@aston.ac.uk} \affiliation{Aston Institute of Photonics
Technologies, Aston University, Aston Triangle, Birmingham, B4 7ET, UK}
\affiliation{Novosibirsk State University, Novosibirsk, 630090 Russia}

\begin{abstract}

Using the path-integral technique we examine the mutual information for the
communication channel modelled  by the nonlinear Schr\"odinger equation with
additive Gaussian noise. The nonlinear Schr\"odinger equation is one of the
fundamental models in nonlinear physics, and it has a broad range of applications,
including fiber optical communications
--- the backbone of the Internet. At large signal-to-noise ratio ($\mathrm{SNR}$) we present
the mutual information through the path-integral which is convenient for the
perturbative expansion in nonlinearity. In the limit of small noise and small
nonlinearity we derive analytically the first nonzero nonlinear correction to the
mutual information for the channel.

\end{abstract}

\maketitle
\section{ Introduction}

There is a known link between entropy production in  physical systems
\cite{Entropy1} and loss of information due to noise in communication channels
\cite{Shannon:1948}. Considering field (signal) evolution in dynamical system with
noise, one can examine a continuous change of the mutual information between the
initial and dynamically evolving fields (signals). The mutual information is a
measure  of the amount of information that can be obtained about one random variable
(in this example - an initial field $X$) by observing another variable (here - the
evolving field $Y$).  The mutual information $I_{P[X]}$  (in continuous-input,
continuous-output system) is expressed  through the path-integral over input $X$ and
output $Y$ fields: \be
\begin{split}
&I_{P[X]}= \int {\cal D}X {\cal D}Y   P[X] P[Y| X]  \log \frac{P[Y|
X]}{P_{out}[Y]}, \label{capacity2}
\end{split}
\ee where $P[X]$ is the probability density function (PDF) of the initial signal $X$
with the fixed finite average power $P_{ave}$. The function $P[Y| X]$ in
Eq.~(\ref{capacity2}) is the conditional probability density function,   that is the
probability density of receiving  output signal $Y$ when  the input signal is $X$.
The output signal PDF $P_{out}[Y]$ in Eq.~(\ref{capacity2}) reads
\begin{align}
& P_{out}[Y]=\int {\cal D}X P[X] P[Y|X]. \label{Pout}
\end{align}
Both signals $X$ and $Y$ may be discrete or continuous. When $X$ is discrete, the
notation integral over $X$ stands for the summation of an under integral function
over its discrete support.  In the traditional communication systems  functions
$X$ and $Y$ usually have a bounded frequency supports, say, the signal $X(\omega)$
is not zero only when $\omega  \in W$ and $Y(\omega)$ is located within the interval $\omega  \in
\widetilde{W}$. In general, domains $W$  and $\widetilde{W}$ might be different, due to
 both nonlinear induced signal spreading in the channel,
and  filtering at the receiver (or inline).

Mutual information (\ref{capacity2}) is a difference between the entropy of the
output signal
\begin{align}
H[Y]=-\int{\cal D}Y P_{out}[Y] \log P_{out}[Y]  \label{HY}
\end{align}
and the conditional entropy
\begin{align}
 H[Y|X]=-\int {\cal D}X {\cal D}Y P[X]P[Y| X] \log P[Y| X]. \label{HYX}
\end{align}
When the signal and noise in the channel are independent variables and the received
signal  $Y$ is the sum of the transmitted signal $X$ and the noise, then it can be
shown explicitly that the entropy of the output signal  $H[Y] $ is greater than the
entropy of the input signal
\begin{eqnarray}
H[X]=-\int{\cal D}X P[X] \log P[X]\,.  \label{HX}
\end{eqnarray}
In this case, the transmission rate is the entropy of the received signal less the
entropy $H[Y|X]$ which is due to impact of the noise. The maximal information
transmission rate over a given bandwidth is given by the maximum of the functional
$I_{P[X]}$  over input field distributions $P[X]$ and is referred to as the channel
(Shannon) capacity $C$. This quantity was calculated for the linear channels with an
additive white Gaussian noise (AWGN) in Ref.~\cite{Shannon:1948}:
\begin{eqnarray}\label{CapacityShannon}
C\propto\log\left(1+\mathrm{SNR}\right)\,,
\end{eqnarray}
where $\mathrm{SNR}$ is a signal-to-noise power ratio. This seminal theoretical
result is the foundation  of communication theory and it has proven its importance
in a number of practical applications. To some extent, the Eq.
(\ref{CapacityShannon}) worked so well in so many situations that some engineers
cease to distinguish the general Shannon expression for capacity and particular
result for the specific linear  channel  with AWGN (\ref{CapacityShannon}).

Recent advances in optical fiber communications where the channel is nonlinear, as
opposite to the linear channel with AWGN, attracted interest to calculation of the
Shannon capacity  for nonlinear channels. To increase the channel capacity over a
certain bandwidth with a  given accumulated noise of optical amplifiers, one has to
increase the signal power, see (\ref{CapacityShannon}). This works in the low
$\mathrm{SNR}$ limit but the refraction index dependence of the fibers on light
intensity (the Kerr effect) dramatically changes the propagation properties of the
fiber optical channel at higher signal power. In other words, the optical fiber
channel becomes nonlinear at high light intensity.

Recent studies have shown that the spectral efficiency (that is, the number of bits,
or nats, transmitted per second per Hertz --- practical characteristics having the
same dimension as the channel capacity per spectral unit) of a fiber optical channel
is limited by the Kerr nonlinearity. These studies indicated that observable
spectral efficiency always turns out to be less  than the Shannon limit of the
corresponding linear channel with AWGN (\ref{CapacityShannon})
\cite{Mitra:2001,Essiambre:2008,Essiambre:2010,Killey:2011,N4}. It has been observed
that  the spectral efficiency of the nonlinear channel decreases with increasing
$\mathrm{SNR}$ at high enough values of $\mathrm{SNR}$
\cite{Mitra:2001,Essiambre:2008,Essiambre:2010,N4}. This analysis certainly provides
only a lower bound on channel capacity and  does not prove that the Shannon
nonlinear fiber channel capacity is decreasing with power; see, for example,
discussions in \cite{Agrell:2012,Agrell:2013,Agrell:2014,tdyt03,TTKhR:2015}. As a
matter of fact, the decrease of the spectral efficiency  can be linked to different
effects. The first effect is the nonlinear interaction of the signal with noise,
which leads to effective increase of the noise power.  The second one is the leak of
the signal power out of the filter domain $\widetilde{W}$ even for  zero noise case,
i.e. not complete collection of the transmitted signal at the receiver.

In \cite{Agrell:2012} it was shown that the capacity of certain nonlinear channels
could not decrease with $\mathrm{SNR}$. Also for the nondispersive nonlinear channel
it was shown that the channel capacity is growing with increasing $\mathrm{SNR}$,
see Refs.~\cite{tdyt03,TTKhR:2015,Mansoor:2011}. However, the capacity of nonlinear
fiber channels is still an open problem of great practical and fundamental
importance. Therefore, it is important to develop new techniques and mathematical
methods to study this problem, especially in the most important case of large
$\mathrm{SNR}$.

In this work we calculate analytically the mutual information for the channel
described by the nonlinear Schr\"odinger equation (NLSE) with AWGN in the leading
nonzero order in nonlinearity and at large $\mathrm{SNR}$. We demonstrate that the
first nonlinear correction for the channel with dispersion is negative and it is
quadratic in the Kerr nonlinearity parameter.  We compare our result for the mutual
information in the case of the channel with nonzero dispersion and the exact result
for the nonlinear nondispersive channel. We show that there is the region of the
parameter $\mathrm{SNR}$ where the obtained mutual information is greater than that
obtained for the channel with  zero dispersion. We also show that the region becomes
wider with increasing of the dispersion parameter.

The article is organized as follows: in the Section \ref{SectionModel} we consider
the channel model and the general structure of the conditional probability density
function. In the Section \ref{SectionEntropy} we obtain the general expressions for
the output signal entropy, conditional entropy and the mutual information. The
Section \ref{SectionCorrection} is focused on the calculation of the first nonlinear
correction to the mutual information and comparison of the result obtained with that
for the nondispersive channel. In the conclusion we discuss our results. The details
of calculation are presented in the Supplementary Materials \cite{SupplMat}.

\section{Nonlinear channel model and the conditional probability at small noise
power } \label{SectionModel}

In our model the propagation of the signal $\psi(z,t)$  is described by the NLSE
with AWGN, see \cite{Iannoe:1998, Turitsyn:2000}:
\begin{eqnarray}\label{startingCannelEqt}
&&\!\!\!\partial_z \psi+i\beta\partial^2_{t}\psi-i\gamma |\psi|^2 \psi=\eta(z,t) \,,
\end{eqnarray}
where $\beta$ is the dispersion coefficient, $\gamma$ is the Kerr nonlinearity
coefficient, $\eta(z,t)$ is an additive complex white noise with zero mean $\langle
\eta (z,t)\rangle_{\eta}=0$ and correlation function
\begin{eqnarray}\label{noisecorrelatort}
\langle \eta (z,t)\bar{\eta}(z^\prime,t^\prime)\rangle_{\eta} =  Q
\delta(z-z^\prime)\delta(t-t^\prime)\,,
\end{eqnarray}
where bar means complex conjugation, and $Q$ is a power of the white Gaussian noise
$\eta(z,t)$  per unit length and per unit frequency. The initial condition for the
signal $\psi(z,t)$ is $\psi(z=0,t)=X(t)$ and we define: $\psi(z=L,t)=Y(t)$. Here $L$
is signal propagation distance. As we mentioned previously  we consider the case
where the input signal $X$ has the bounded frequency support $W$. Therefore it is
convenient to consider the problem in the frequency domain. Any functions in the
time and frequency domains are related as follows: $f(z,t)=\int \frac{d\omega}{2\pi}
e^{-i \omega t} f_{\omega}(z)$. In the frequency domain our Eqs.
(\ref{startingCannelEqt}) and (\ref{noisecorrelatort}) have the form:
\begin{eqnarray}\label{startingCannelEq}
&&\partial_z \psi_\omega (z)-i\beta\omega^2\psi_\omega (z)-\nonumber \\&&
i\gamma\int^{\infty}_{-\infty}\frac{d\omega_1d\omega_2}{(2\pi)^2}
 \psi_{\omega_1} (z) \psi_{\omega_2} (z)
\bar{\psi}_{\omega_3} (z)= \eta_\omega (z) \,,
\end{eqnarray}
where $\omega_3= \omega_1+\omega_2-\omega$,
\begin{eqnarray}\label{noisecorrelatoromega}
\langle \eta_\omega (z)\bar{\eta}_{\omega^\prime} (z^\prime)\rangle_{\eta} = 2\pi Q
\delta(z-z^\prime)\delta(\omega-\omega^\prime)\chi_{W'}(\omega)\,,
\end{eqnarray}
where $\chi_{W'}(\omega)=\theta(W'/2-\omega)\theta(W'/2+\omega)$, with $\theta(x)$
being Heaviside $\theta$-function. Strictly speaking the finite frequency domain of
the noise means that the noise is not white having the finite frequency support
$W'$. But if $W'$ is much larger than the frequency domain of the signal
$\psi_\omega (z)$ (i.e. $W' \gg W$ and $W' \gg \widetilde{W}$) then the noise can be
treated as a white one. Our results will not depend on the parameter $W'$ and at the
final stage we consider infinitely large $W'$ (true white noise). It is worth
emphasizing that in a nonlinear channel transmitted and received signal bandwidths
can differ from each other. Therefore,  we assume here that in general, the input
$X(\omega)$  and output $Y(\omega)$ signals have frequency domains $[-W/2,\,W/2]$
and $[-\widetilde{W}/2,\,\widetilde{W}/2]$ respectively.


{\it The model of the input signal $X$.} We imply that  the input signal $X(\omega)$
is not zero in the frequency domain $W$ and $X(\omega)=0$ in the domain $W'
\setminus W$. In the  domain $W$ the signal $X(\omega)$ has the PDF with zero mean
and with fixed average power. Since $X(\omega)$ in the  domain $W' \setminus W$ is
defined and is equals to zero the PDF in the domain has the form of delta-function.
Therefore one has
\begin{eqnarray}
&\mspace{-20mu}P[X(\omega)]=  P^{(M)}_{X}[X(\omega)] \prod^{M'-M}_{j \in W'
\setminus W} \mspace{-2mu} \delta( X_{j})\,.\label{PX}
\end{eqnarray}
Here we divide the domain $W'$ into $M'$ equal intervals and the domain $W$ into $M$
equal intervals. The form (\ref{PX}) stands for the fact that we have $M$
independent complex meaning channels in the domain $W$ with the same PDF in every
channel:
\begin{eqnarray} \label{PXM}
&P^{(M)}_{X}[X(\omega)]=  \prod^{M}_{j=1}P[X_{j}].
\end{eqnarray}
Here $\delta(X_j)=\delta( Re\, X_j) \delta(Im\, X_j)$ is  the $\delta$-function,
$X_{j}=X(\omega_j)$. The frequency domain $W$ ($W'$) is divided by $M$ ($M'$) grids
spacing $\delta={W}/{(2\pi M)} = {W'}/{(2\pi M')}$. The distribution (\ref{PX})
means that there are  $M$  elementary independent complex coefficients  presenting
information in the spectral domain $W$. The average power for $P[X(\omega)]$ reads
\begin{eqnarray}\label{Pave}
P_{ave}=\lim_{T\to\infty}\!\!\int \!\!{\cal D}X P[X(\omega)] \int_{W'}
\frac{d\omega}{2\pi T} |X({\omega})|^2=  \frac{P W}{2\pi}\,,
\end{eqnarray}
where $T$ is the time interval containing the whole input signal in the time domain.
We will use the relation $M=T W/2\pi$ that corresponds to the
Nyquist-Shannon-Kotelnikov theorem \cite{Shannon:1949}. In Eq.~(\ref{Pave}) we have
introduced quantity $P$ that is the  power per unit frequency (spectral power
density), it means that the average power in one elementary spectral step is $P
\delta$. The measure in Eq.~(\ref{Pave}) ${\cal D}X =\prod_{j=1}^{M'} d Re\, X_j \,
d Im \, X_j$ is consistent with the normalization condition
\begin{eqnarray}\label{PXnorm}
\int {\cal D}X P[X(\omega)] =1.
\end{eqnarray}
For typical fiber optical links the ratio $\mathrm{SNR}=P/(QL)$ is of order of
$10^4$. Therefore, in what follows we assume that the parameter $P$ is much greater
than the accumulated noise power $QL$ in the channel (large $\mathrm{SNR}$ case):
\begin{eqnarray}\label{PandQ}
P \gg QL.
\end{eqnarray}

To calculate the conditional  probability density $P[Y(\omega)| X(\omega)]$ we use
the technique described in Ref. \cite{Terekhov:2014}. In this formalism the function
$P[Y(\omega)| X(\omega)]$ can be expressed through the path-integral:

\begin{eqnarray} \label{PYXquasiclassic}
P[Y(\omega)|X(\omega)]=\Lambda \, e^{-{S[\Psi_{\omega}(z)]}/{Q}},
\end{eqnarray}
\begin{eqnarray}
&& \Lambda=\int^{\phi_{\omega}(L)=0}_{\phi_{\omega}(0)=0} \!\!\!\!{\cal D} \phi\,
e^{ -\left\{S[\Psi_{\omega}(z)
+\phi_{\omega}(z)]-S[\Psi_{\omega}(z)]\right\}/Q}\,.\label{PYXquasiclassicNorm}
\end{eqnarray}
Here the functional $S[\psi]$ is referred to as the action, and it has the form
\begin{eqnarray}
&&S[\psi]=\int^L_0 dz \int_{W'} \frac{d\omega}{2\pi} \Big|\partial_z
\psi_{\omega}(z)-i \beta \omega^2 \psi_{\omega}(z) - \nonumber \\&&
i\gamma\int_{W'}\frac{d\omega_1d\omega_2}{(2\pi)^2} \psi_{\omega_1} (z)
\psi_{\omega_2} (z) \bar{\psi}_{\omega_3} (z) \Big|^2\,,\label{Action}
\end{eqnarray}
where $\omega_3= \omega_1+\omega_2-\omega$. The function $\Psi_{\omega}(z)$ in
Eq.~(\ref{PYXquasiclassic}) is referred to as the ``classical trajectory''. It is
the extremum function of the action $S$, i.e.  the action variation is equal to zero
on the function $\Psi_{\omega}(z)$: $\delta S[\Psi]=0$ with  the boundary conditions
$\Psi_{\omega}(0)=X(\omega)$, $\Psi_{\omega}(L)=Y(\omega)$. We omit here the
explicit form of the equation because it is quite cumbersome, but one can find it in
\cite{SupplMat}. The path-integral in Eq.~(\ref{PYXquasiclassicNorm}) is defined in
the discretization scheme that takes into account the casuality principle, see
details in Ref.~\cite{Terekhov:2014}. The measure ${\cal D} {\phi}$ in
Eq.~(\ref{PYXquasiclassic}) is defined as
\begin{eqnarray} \label{measureDphi}
{\cal D} {\phi}=\lim_{\delta \rightarrow 0}\lim_{\Delta \rightarrow 0} \Big(
\frac{\delta}{\Delta \pi Q}\Big)^{N
M'}\prod^{M'}_{j=1}\prod^{N-1}_{i=1}d{\phi}_{i,\,j},
\end{eqnarray}
where $d{\phi}_{i,\,j}=d\,Re{\phi}_{i,\,j}\,d\,Im {\phi}_{i,\,j}$,
${\phi}_{i,\,j}={\phi}_{\omega_j}(z_i)$,  $\Delta={L}/{N}$ is the coordinate grids
spacing, $\delta$ is the frequency grids spacing introduced after Eq.~(\ref{PXM}).
The measure (\ref{measureDphi}) is consistent with the normalization condition
\begin{eqnarray} \label{PYXnorm}
\int {\cal D} {Y} P[Y(\omega)|X(\omega)]=1,
\end{eqnarray}
where the measure ${\cal D} {Y}$ is defined as
\begin{eqnarray} \label{measureDY}
{\cal D} {Y}= \prod_{j=1}^{M'} d \,Re\, Y_j \,\, d \,Im \, Y_j, \qquad
Y_j=Y(\omega_j).
\end{eqnarray}

Let us now consider the function $P[Y(\omega)|X(\omega)]$ at small $Q$. Our
consideration of the $P[Y|X]$ at small parameter $Q$ is similar to the
quasi-classical approximation in the quantum mechanics at small Planck's constant
$\hbar$ \cite{Landau:1981}. Let us consider what output signals $Y(\omega)$ are
statistically significant for $P[Y(\omega)|X(\omega)]$ at given $X(\omega)$, i.e.
when $S[\Psi]$ is less or of order of $Q$. The physical picture is as follows. At
small $Q$ the trajectory $\Psi_{\omega}(z)$ can not be sufficiently different from
$\Phi_{\omega}(z)$ which is the  solution of Eq.~ (\ref{startingCannelEq}) with zero
noise $\eta=0$ and with the boundary condition $\Phi_{\omega}(0)=X(\omega)$. When
solving (\ref{startingCannelEq}) with the fixed $X(\omega)$ and nonzero (but small)
noise $\eta$ we can expect that the solution at $z=L$, $\psi_{\omega}(L)$, has the
difference from $\Phi_{\omega}(z=L)$ proportional to $\sqrt{Q}$ because the average
noise power per unit frequency is small, see Eq.~(\ref{noisecorrelatoromega}).  That
is why the difference $Y(\omega)-\Phi_{\omega}(L)$ should be proportional to
$\sqrt{QL}$. Thus, for such $Y(\omega)$ we can seek the solution $\Psi_{\omega}(z)$
as the series in parameter $\sqrt{QL}$:
\begin{eqnarray} \label{Psitokappan}
\Psi_{\omega}(z)=\Phi_{\omega}(z)+\varkappa_{\omega}(z),\qquad
\varkappa_{\omega}(z)=\sum^{\infty}_{n=1}\varkappa^{(n)}_{\omega}(z),
\end{eqnarray}
where $\varkappa^{(n)}_{\omega}(z) \propto (QL)^{n/2}$. Inserting the solution
(\ref{Psitokappan}) in the action $S[\Psi]$ and taking into account that $S[\Phi]=0$
we obtain $S[\Psi]=S_2\left[\varkappa^{(1)}\right]+\widetilde{S}[\varkappa]$, where
$S_2\left[\varkappa^{(1)}\right]$ is the quadratic functional in $\varkappa^{(1)}$,
i.e. $S_2\left[\varkappa^{(1)}\right] \propto Q$, and $\widetilde{S}[\varkappa]$ is
the reminder functional that is suppressed in the parameter $Q$ (its expansion in
$Q$ starts from $Q^{3/2}$). In what follows we are interested only in  the leading
order in parameter $QL$, therefore:
\begin{eqnarray} \label{S2kappa}
&&S[\Psi] \approx S_2\left[\varkappa^{(1)}\right]= \int^L_0 dz \int_{W'}
\frac{d\omega}{2\pi} \Big|\partial_z \varkappa^{(1)}_{\omega}-i \beta \omega^2
\varkappa^{(1)}_{\omega} - \nonumber \\&& \!\!\!i \gamma \int_{W'}
\frac{d\omega_1d\omega_2}{(2\pi)^2} \left( 2
\varkappa^{(1)}_{\omega_1}\Phi_{\omega_2} \bar{\Phi}_{\omega_3} +
\bar{\varkappa}^{(1)}_{\omega_3}\Phi_{\omega_1} {\Phi}_{\omega_2}\right)\Big|^2,
\end{eqnarray}
where $\omega_3=\omega_1+\omega_2-\omega$. The function
$\varkappa^{(1)}_{\omega}(z)$ obeys the linear equation with coefficients depending
on $\Phi_{\omega}(z) $ with the boundary conditions $\varkappa^{(1)}_{\omega}(0)=0$,
$\varkappa^{(1)}_{\omega}(L)=Y(\omega)-\Phi_{\omega}(L)$. The equation for
$\varkappa^{(1)}$ has a compact form in the time domain:
\begin{eqnarray}\label{kappaequation}
&&\!\!\!\left(\partial_z+i\beta\partial^2_t -2 i \gamma |\Phi|^2\right)
l[\varkappa^{(1)}]+i\gamma \Phi^2\, \bar{l}[\varkappa^{(1)}]=0,  \\
&& l[\varkappa]=\left(\partial_z+i\beta\partial^2_t\right)\varkappa- i \gamma
\left(2|\Phi|^2 \varkappa+ \Phi^2 \overline{\varkappa}\right).
\end{eqnarray}
This equation in the frequency domain is cumbersome, therefore, we do not present it
here but one can find it in \cite{SupplMat}. Since the Eq.~(\ref{kappaequation}) is
linear in $\varkappa^{(1)}$ the solution of the equation for
$\varkappa^{(1)}_{\omega}(z)$ linearly depends on its value $\delta
Y(\omega)=Y(\omega)-\Phi_{\omega}(L)$ on the boundary $z=L$. Since the action
(\ref{S2kappa}) is quadratic functional in $\varkappa^{(1)}_{\omega}(z)$ we can
write
\begin{eqnarray}
\label{Lomegaomega} \!\!\!S[\Psi]\approx\int d\omega\,d\omega'\delta
{Y}^{(\alpha)}(\omega){\cal L}_{\alpha,\,\beta}(\omega,\omega')\delta
Y^{(\beta)}(\omega'),
\end{eqnarray}
where $\delta Y^{(1)}(\omega)= Re\, \delta Y(\omega)$, $\delta Y^{(2)}(\omega)= Im\,
\delta Y(\omega)$, and ${\cal L}_{\alpha,\,\beta}(\omega,\omega')$, ($\alpha, \,
\beta=1,\,2$) is some integral kernel that depends on function $\Phi_{\omega}(z)$.
Note that the solution $\Phi_{\omega}(z)$ of Eq.~(\ref{startingCannelEq}) can be
written as $\Phi_{\omega}(L)= \left(\hat{L} X \right)(L,\omega)\equiv \hat{L} X$,
where $\hat{L}$ is  the nonlinear evolution operator of
Eq.~(\ref{startingCannelEq}), see \cite{Zakharov:1984}. It means that in the leading
order in $Q$ the kernel ${\cal L}_{\alpha,\,\beta}(\omega,\omega')$ depends on input
signal $X(\omega)$ rather than $Y(\omega)$. The representation (\ref{Lomegaomega})
is valid for arbitrary nonlinearity but in the leading order in $Q$.

Let us consider the normalization factor $\Lambda$ in
Eq.~(\ref{PYXquasiclassicNorm}). In order to calculate $\Lambda$ in the  leading
order in $Q$ we should keep only the quadratic in $\phi_{\omega}(z)$ terms in the
action difference, see Eq.~(\ref{PYXquasiclassicNorm}). Using Laplace's method
applied to the path-integral one can show that the higher powers of
$\phi_{\omega}(z)$ result in the suppressed corrections in the parameter $Q$. The
coefficients in the quadratic in $\phi_{\omega}(z)$ terms in the action difference
depend on the function $\Psi$, but in the leading order in $Q$ we can substitute
$\Phi$ instead of $\Psi$.  It means that in this order the normalization factor
$\Lambda=\Lambda[X]$ depends only on $X(\omega)$. In the leading order in $Q$ the
factor $\Lambda[X]$ can be found in several ways: by the direct calculation of the
path-integral or by using the normalization condition (\ref{PYXnorm}). The latter
reads
\begin{eqnarray} \label{normPYX}
\int {\cal D} Y P[Y|X]= \Lambda[X] \int {\cal D} \delta Y
e^{-{S[\Psi_{\omega}(z)]}/{Q}} =1.
\end{eqnarray}
In the discrete form the functions $X(\omega)$ and $Y(\omega)$ can be presented as
$2M'$-dimensional real vectors $\vec{X}$ and $\vec{Y}$, respectively,  which
describe both real and imaginary part of these quantities on the frequency grid.
Thus Eq.~(\ref{Lomegaomega}) reads in the discretization as follows:
\begin{eqnarray} \label{Actiondiscrete}
S[\Psi]\approx \delta^2 \vec{\delta {Y}}^{\dagger} {\cal L} \, \vec{\delta {Y}},
\end{eqnarray}
with ${\cal L}={\cal L}\left[\vec{X}\right]$ being $2 M' \times 2 M'$-dimensional
Hermitian matrix depending on $\vec{X}$ only. For $\Lambda[X]$ one has
\begin{eqnarray} \label{normPYX2}
\Lambda[X] =\sqrt{\det[{\cal L}] }\left({\delta^{2}}/{(\pi Q)}\right)^{M'}.
\end{eqnarray}
Therefore in the leading order in $Q$ the conditional probability density function
$P[Y|X]$ has the form:
\begin{eqnarray} \label{PYX}
P[Y|X]=\Lambda[X]\, e^{-\delta^2 \vec{\delta {Y}}^{\dagger} {\cal L} \, \vec{\delta
{Y}}/Q}.
\end{eqnarray}
The conditional probability density function $P[Y|X]$ must obey the restriction
\cite{TTKhR:2015,Terekhov:2014}:
\begin{eqnarray} \label{PYXQzero}
\lim_{Q \to 0}P[Y|X]=\delta(\vec{Y}-\overrightarrow{\hat{L} X}),
\end{eqnarray}
that is nothing more but the deterministic limit of zero noise. In our approximation
for the $P[Y|X]$ this condition (\ref{PYXQzero}) is fulfilled automatically due to
the exponential form (\ref{PYX}) and normalization factor (\ref{normPYX2}). Now we
can move to the consideration of the output and conditional  signal entropies
(\ref{HY}) and  (\ref{HYX}), respectively.

\section{Entropies and mutual information}
\label{SectionEntropy}

First we consider the PDF $P_{out}[Y]$, see Eq.~ (\ref{Pout}). To begin with we
perform the decomposition of any $2M'$-vector ($\vec{X}$, $\vec{Y}$, etc.) ${\vec
V}={\vec V}_1 \oplus {\vec V}_2$, where ${\vec V}_1$ is $2M$-dimensional vector
corresponding to $M$ meaning complex channels in the frequency domain $W$, whereas
${\vec V}_2$ is $2(M'-M)$-dimensional vector corresponding to remnant  $M'-M$
complex channels in the frequency domain $W'\setminus W$. The sign $\oplus$ means
the direct sum. We substitute the PDF $P[X(\omega)]$ in the form (\ref{PX}) and the
conditional PDF $P[Y|X]$ in the form (\ref{PYX}) into the definition~(\ref{Pout})
and obtain $P_{out}[\vec{Y}]$ in the discretization scheme:
\begin{eqnarray} \label{Poutdiscrete}
P_{out}[\vec{Y}]&=&\int {d }\vec{X}_{1} {d}\vec{X}_{2}P^{(M)}_{X}[\vec{X}_1]
\delta\left(\vec{X}_2\right) \times \nonumber \\&& \Lambda[\vec{X}]\, e^{-\delta^2
\vec{\delta {Y}}^{\dagger} {\cal L} \, \vec{\delta {Y}}/Q},
\end{eqnarray}
where $\vec{\delta {Y}}=\vec{Y}-\overrightarrow{\hat{L} X}$,
$\delta\left(\vec{X}_2\right)$ means $2(M'-M)$-dimensional delta-function. For the
following calculation it is convenient to perform the transformation of the action
(\ref{Actiondiscrete}). We can write $\vec{X}=\overrightarrow{\hat{L}^{-1} Y} -J
\vec{\delta {Y}}+{\cal O}(\vec{\delta {Y}}^2)$ using that $\vec{\delta {Y}} \sim
\sqrt{Q}$. Here $J_{i,\,i'}={\partial \hat{L}^{-1} Y _{i}}/{\partial Y_{i'}}$ is the
Jacobian matrix of the mapping $\hat{L}^{-1}$, $i,\,i'=1,\ldots, 2M'$. Since the
Jacobian $\det[J]$ has the unit absolute value, see \cite{Zakharov:1984}, we can
write $\vec{\delta {Y}}=-J^{-1} \vec{Z}$, where
$\vec{Z}=\vec{X}-\overrightarrow{\hat{L}^{-1} Y}$. Now we change variables in
Eq.~(\ref{Poutdiscrete}) from $\vec{X}$ to  $\vec{Z}=\vec{Z}_1 \oplus \vec{Z}_2$. In
the new variables the action (\ref{Actiondiscrete}) reads $S[\Psi]\approx \delta^2
\vec{Z}^{\,\dagger} {\cal K} \, \vec{Z}$, where Hermitian matrix  ${\cal
K}=J^{-1\,\dagger}{\cal L}\,J^{-1}$ has the block form
\begin{eqnarray}\label{Kmatrix}
&& {\cal K}=J^{-1\,\dagger}{\cal L}\,J^{-1}=\begin{pmatrix}
{\cal K}_{1\,1}, & {\cal K}_{1\,2}&  \\
{\cal K}_{2\,1}, & {\cal K}_{2\,2}&
\end{pmatrix}.
\end{eqnarray}
Here the block ${\cal K}_{1\,1}$ is $2M \times 2M$ matrix, ${\cal K}_{1\,2}$ is $2M
\times 2(M'-M)$ matrix, ${\cal K}_{1\,2}={\cal K}^{\dagger}_{2\,1}$, the block
${\cal K}_{2\,2}$ is $ 2(M'-M) \times  2(M'-M)$ matrix. In the new variables the
action (\ref{Actiondiscrete}) has the form
\begin{eqnarray}\label{asctionZ}
S[\Psi]\!\!&\approx&  \!\!\left(\vec{Z}_1+{\cal K}^{-1}_{1\,1} {\cal K}_{1\,2}
\vec{Z}_2\right)^{\dagger}\!{\cal K}_{1\,1}\!\left(\vec{Z}_1+{\cal K}^{-1}_{1\,1}
{\cal K}_{1\,2} \vec{Z}_2\right)\delta^2+ \nonumber \\&&
\vec{Z}^{\,\dagger}_2\left({\cal K}_{2\,2}- {\cal K}_{2\,1} {\cal K}^{-1}_{1\,1}
{\cal K}_{1\,2}\right)\vec{Z}_2\, \delta^2,
\end{eqnarray}
where matrices ${\cal K}_{\alpha\,\beta}$ depend on
$\vec{X}=\vec{Z}+\overrightarrow{\hat{L}^{-1} Y}$. Now we can perform the
integration over $d \vec{Z}_2$, everywhere substituting
$-\overrightarrow{\hat{L}^{-1} Y}_2$ instead of $\vec{Z}_2$  in view of the
delta-function. Passing to new variables $\vec{Z}'_1=\vec{Z}_1-{\cal K}^{-1}_{1\,1}
{\cal K}_{1\,2}\overrightarrow{\hat{L}^{-1} Y}_2$ we obtain in the leading order in
$1/\mathrm{SNR}$
\begin{eqnarray} \label{Poutdiscrete1}
&&\!\!\!P_{out}[\vec{Y}]=\!\int \!{d }\vec{Z}'_{1}
P^{(M)}_{X}[\vec{Z}'_{1}+\overrightarrow{\hat{L}^{-1} Y}_1+{\cal K}^{-1}_{1\,1}
{\cal K}_{1\,2}\overrightarrow{\hat{L}^{-1} Y}_2] \times \nonumber \\&&
\Lambda[\vec{Z}'_{1}+\overrightarrow{\hat{L}^{-1} Y}_1+{\cal K}^{-1}_{1\,1} {\cal
K}_{1\,2}\overrightarrow{\hat{L}^{-1} Y}_2] e^{-\delta^2 \vec{ {Z}}'^{\dagger}_1
{\cal K}_{1\,1} \, \vec{ {Z}}'_1/Q} \times \nonumber \\&&
\exp\left[-\frac{\delta^2}{Q}\overrightarrow{\hat{L}^{-1}
Y}_2^{\,\dagger}\left({\cal K}_{2\,2}- {\cal K}_{2\,1} {\cal K}^{-1}_{1\,1} {\cal
K}_{1\,2}\right)\overrightarrow{\hat{L}^{-1} Y}_2\right].
\end{eqnarray}
Since the first exponent $e^{-\delta^2 \vec{ {Z}}'^{\dagger}_1 {\cal K}_{1\,1} \,
\vec{ {Z}}'_1/Q}$ in Eq.~(\ref{Poutdiscrete1}) is essentially narrower than the
function $P^{(M)}_{X}[X]$, see Eq.~(\ref{PandQ}), we can set $\vec{ {Z}}'_1=0$ in
the argument of $P^{(M)}_{X}[X]$ and in the argument of $\Lambda[X]$. The second
exponent in Eq.~(\ref{Poutdiscrete1}) demonstrates that
$\overrightarrow{\hat{L}^{-1} Y}_2 \propto \sqrt{Q}$ as well, therefore in the
leading order in $1/\mathrm{SNR}$ we can omit $\overrightarrow{\hat{L}^{-1} Y}_2$ in
the arguments of $P^{(M)}_{X}[X]$ and $\Lambda[X]$. After this simplifications we
perform the Gaussian integration over $\vec{ {Z}}'_1$ and finally obtain
\begin{eqnarray} \label{Poutdiscrete2}
&&P_{out}[\vec{Y}]=P^{(M)}_{X}[\overrightarrow{\hat{L}^{-1} Y}_1] \times \nonumber
\\&& \Lambda_2 \, e^{-\delta^2 \overrightarrow{\hat{L}^{-1} Y}^{\dagger}_2 \left(
{\cal K}_{2\,2}- {\cal K}_{2\,1} {\cal K}^{-1}_{1\,1} {\cal K}_{1\,2} \right)
\overrightarrow{\hat{L}^{-1} Y}_2/Q},
\end{eqnarray}
where
\begin{eqnarray}
\label{Lambda2} \Lambda_{2}=\sqrt{\det[{\cal K}_{2\,2}- {\cal K}_{2\,1} {\cal
K}^{-1}_{1\,1} {\cal K}_{1\,2}] }\left({\delta^{2}}/{(\pi Q)}\right)^{M'-M}.
\end{eqnarray}
Here the matrices ${\cal K}_{\alpha\,\beta}$ depend on vector
$\overrightarrow{\hat{L}^{-1} Y}_1$. To obtain Eq.~ (\ref{Poutdiscrete2}) we have
used the  factorization identity
\begin{eqnarray} \label{Lambdarelation}
&&\Lambda[X] = \Lambda_{1}[X]\times \Lambda_{2}[X],
\end{eqnarray}
where $\Lambda[X]$ is given by Eq.~(\ref{normPYX2}), i.e. $\Lambda[X]
=\sqrt{\det[{\cal L}] }\left({\delta^{2}}/{(\pi Q)}\right)^{M'}=\sqrt{\det[{\cal K}]
}\left({\delta^{2}}/{(\pi Q)}\right)^{M'}$, and
\begin{eqnarray} \label{Lambda1}
&& \Lambda_{1}[X]=\sqrt{\det[{\cal K}_{1\,1}] }\left({\delta^{2}}/{(\pi
Q)}\right)^{M}.
\end{eqnarray}
Let us note that  $P_{out}[\vec{Y}]$ obtained in the leading order in
$1/\mathrm{SNR}$, see Eq.~(\ref{Poutdiscrete2}),  is a product of the initial signal
PDF $P^{(M)}_{X}$ in $M$ complex meaning channels and  some noise distribution in
the  other $M'-M$ complex channels which depends on the signal
$\overrightarrow{\hat{L}^{-1} Y}_1$ of the meaning channels through the matrices
${\cal K}_{\alpha\,\beta}$.

Now we can calculate the output signal entropy $H[Y]$, see Eq.~(\ref{HY}). To this
end we insert  $P_{out}$ in the form (\ref{Poutdiscrete2}) to the Eq.~(\ref{HY}),
then change the integration variables from $\vec{Y}$ to $\vec{N}=
\overrightarrow{\hat{L}^{-1} Y}$. Next using the fact that the Jacobian $\det[J]$
has the unit absolute value we perform integration over $\vec{N}_2$ and  obtain:
\begin{eqnarray} \label{HY2}
&& H[Y]=H[X]+(M'-M)-\nonumber \\&&\int d\vec{N}_1  P^{(M)}_{X}[\vec{N}_1 ]\,\log
\Lambda_2[\vec{N}_1],
\end{eqnarray}
where
\begin{eqnarray} \label{HXM}
&& H[X]=-\int d\vec{N}_1 P^{(M)}_{X}[\vec{N}_1 ] \log P^{(M)}_{X}[\vec{N}_1 ]
\end{eqnarray}
is the entropy of the input signal $X$, see Eq.~(\ref{HX}).

Next, we  calculate the conditional entropy $H[Y|X]$, see Eq. (\ref{HYX}). This
calculation is similar to the one performed above. First, we perform the integration
over $\vec{X}_2$. Then we change the variables $\vec{X}_1$ to
$\vec{Z}'_1=\vec{X}_1-\overrightarrow{\hat{L}^{-1} Y}_1-{\cal K}^{-1}_{1\,1} {\cal
K}_{1\,2}\overrightarrow{\hat{L}^{-1} Y}_2$. Then we change the variables $\vec{Y}$
to $\vec{N}= \overrightarrow{\hat{L}^{-1} Y}$. After that we perform integration
over $\vec{N}$ and then over $\vec{Z}'_1$. Finally, we obtain the conditional
entropy $H[Y|X]$ in the leading order in $1/\mathrm{SNR}$:
\begin{eqnarray} \label{HYXresult}
H[Y|X]&=&M'-\int d\vec{Z}'_1  P^{(M)}_{X}[\vec{Z}'_1 ]\,\log \Lambda_1[\vec{Z}'_1 ]-
\nonumber \\&& \int d\vec{Z}'_1 P^{(M)}_{X}[\vec{Z}'_1]\,\log \Lambda_2[\vec{Z}'_1].
\end{eqnarray}

To obtain the mutual information (\ref{capacity2}) we subtract $H[Y|X]$, see Eq.~
(\ref{HYXresult}), from $H[Y]$, see Eq.~ (\ref{HY2}), and get
\begin{eqnarray} \label{Mutualinformation1}
I_{P[X]}= H[X]-M+\int d\vec{Z}'_1  P^{(M)}_{X}[\vec{Z}'_1 ]\,\log
\Lambda_1[\vec{Z}'_1 ].
\end{eqnarray}
Note that the mutual information $I_{P[X]}$ depends only on $M$ complex
coefficients,  whereas the entropies (\ref{HY2}) and (\ref{HYXresult})  depend on
$M'$ complex parameters. One can see that in the leading order in $1/\mathrm{SNR}$
our result (\ref{Mutualinformation1}) contains the initial signal entropy $H[X]$ and
the logarithm of the normalization factor $\Lambda_1$ averaged over the initial
signal distribution $P^{(M)}_{X}$. Therefore to calculate $I_{P[X]}$ we have to know
the normalization factor $\Lambda_1$.

\section{First nonlinear correction}
\label{SectionCorrection}

In this  section we consider the mutual information (\ref{Mutualinformation1}) in
different regimes in the case when the input signal PDF $P^{(M)}_{X}$ has the
Gaussian form. First, we examine the mutual information in the limit of small
nonlinearity: when the dimensionless parameter $\tilde{\gamma}=P_{ave}{\gamma}L$ is
small.  In this case we calculate the first nonzero nonlinear correction to the
mutual information for the arbitrary dispersion parameter $\beta$. Secondly, we
consider the mutual information for the arbitrary nonlinearity and zero dispersion.

To find the mutual information (\ref{Mutualinformation1}) at small $\tilde{\gamma}$
we should calculate the initial signal entropy $H[X]$, see Eq.~(\ref{HXM}), and the
normalization factor $\Lambda_1$. We use
the Gaussian input signal PDF $P^{(M)}_{X}$ in the form
\begin{eqnarray}
P^{(M)}_{X}[\vec{X}_1]=P_{G}[\vec{X}_1]=\Lambda_{P} \, e^{- |\vec{X}_1|^2
\delta/P},\, \label{PXGaussian}
\end{eqnarray}
where $\Lambda_{P}$ is consistent with the normalization condition (\ref{PXnorm})
and has the form:
\begin{eqnarray}
\Lambda_{P}=({\delta}/{(\pi P)})^{M}\label{LambdaGauss}.
\end{eqnarray}
The input signal PDF in the form (\ref{PXGaussian}) means that the average signal
power (\ref{Pave}) is $P_{ave}=P W/(2\pi) \gg P_{noise}=QL W/(2\pi)$. The
normalization condition reads $\int {d}\vec{X}_1 P_{G}[\vec{X}_1] =1$. Substitution of
the PDF (\ref{PXGaussian}) into Eq.~(\ref{HXM}) and the following integration yields:
\begin{eqnarray} \label{HXGaussian}
&&H[X]= M+M\log (\pi P/\delta).
\end{eqnarray}

\begin{figure}[h]
\begin{center}
\includegraphics[width=6cm]{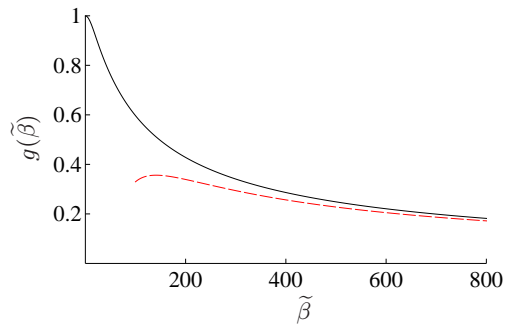}
\begin{picture}(0,0)
\put(-82,-10){\text{$\widetilde{\beta}$}}
\put(-190,50){\rotatebox{90}{\text{$g(\widetilde{\beta})$}}}
\end{picture}
\end{center}
\caption{\label{fig1} The function $g(\widetilde{\beta})$. The solid black line
corresponds to exact expression (\ref{g-function4}) for $g(\widetilde{\beta})$, the
red dashed line corresponds to the asymptotics (\ref{g-functionlarge}) of
$g(\widetilde{\beta})$ at large $\widetilde{\beta}$.}
\end{figure}

To calculate the averaged  $\log \Lambda_1$ over PDF $P_{G}[\vec{X}_1]$ in
Eq.~(\ref{Mutualinformation1}) in the leading and next-to-leading order in
$\tilde{\gamma}$ we have to factorize $\Lambda$ in the form of the path-integral
Eq.~(\ref{PYXquasiclassicNorm}), see Eq.~(\ref{Lambdarelation}). We divide the
integration region  $W'$ of variable $\omega$ in the action Eq.~(\ref{Action}) into
two subregions $W$ and $W' \setminus W$. The first subregion $W$ results in the
normalization factor  $\Lambda_1$ whereas the subregion $W' \setminus W$ results in
the factor $\Lambda_2$. Despite the nonlinearity term in the action ~(\ref{Action})
the fields $\phi_{\omega}$ for the different subregions do not mix in these orders
in $\tilde{\gamma}$. Therefore $\Lambda$ can be expressed as the product of two
path-integrals. The first integral contains the fields $\phi_{\omega}$ for $\omega$
from the subregion $W$ and corresponds to $\Lambda_1$. The second one corresponds to
$\Lambda_2$. Therefore the normalization factor $\Lambda_1$ can be expressed in the
form of the path-integral (\ref{PYXquasiclassicNorm}) where all frequencies are from
$W$ subregion. The details of the factorization of $\Lambda$ and the details of
calculation of the averaged $\log \Lambda_1$ over PDF $P_{G}[\vec{X}_1]$ are
presented  in Ref.~\cite{SupplMat}. Here we present only the final result:
\begin{eqnarray}
I_{P_G[X]}=M \log \mathrm{SNR}- M \frac{\widetilde{\gamma}^2}{3}
g(\widetilde{\beta})+{\cal O}(\widetilde{\gamma}^{\,4}), \label{MutInfRes}
\end{eqnarray}
where $g(\widetilde{\beta})$ is the function of dimensionless parameter
$\widetilde{\beta}=\beta L W^2$:
\begin{eqnarray}\label{g-function4}
\!\!\!\!\!\!\!g(\widetilde{\beta})=4!\sum^{\infty}_{n=0} \frac{(-1)^n
\widetilde{\beta}^{2n} \left[(4n+2)! +
(1+2n)!^2\right]}{2^{2n-1}(2n+4)!(4n+3)!(1+2n)^2}.
\end{eqnarray}
One can check that $g(\widetilde{\beta}=0)=1$. In the case when  $\widetilde{\beta}
\gg 1$ the asymptotics for the function (\ref{g-function4}) reads
\begin{eqnarray}\label{g-functionlarge}
&&\!\!\!\!\!\!\!\!g(\widetilde{\beta})\sim \frac{16 \pi}{\widetilde{\beta}}\left(
\log \frac{\widetilde{\beta}}{2} +\gamma_{E} -\frac{23}{6}\right)+{\cal O}\left(
{\widetilde{\beta}^{-3/2}}\right),
\end{eqnarray}
where $\gamma_{E} \approx 0. 577$ is the Euler constant. The function
$g(\widetilde{\beta})$ is plotted in Fig.~\ref{fig1}.
Note that the result (\ref{MutInfRes}) is proportional to the number of the meaning
channels $M$. The reason for that is the definition of the mutual information
through the path-integral (\ref{capacity2}). Usually instead of the mutual
information (\ref{capacity2}) the spectral efficiency is considered as the quantity
which does not depend on $M$:
\begin{eqnarray}
i_{P[X]}&=&\lim_{T \to \infty}\frac{2\pi}{ T W}I_{P[X]}= \frac{I_{P[X]}}{M}=
\nonumber \\&& \log \mathrm{SNR}-  \frac{\widetilde{\gamma}^2}{3}
g(\widetilde{\beta})+{\cal O}(\widetilde{\gamma}^{\,4}), \label{SE}
\end{eqnarray}
where the parameter $T$ is the time duration of the signal. The quantity $i_{P[X]}$
coincides with the per-sample mutual information for the nondispersive case
$\beta=0$.


Let us consider the mutual information (\ref{Mutualinformation1}) at zero $\beta$.
For the nondispersive case the result for the per-sample mutual information was
obtained in Ref.~\cite{TTKhR:2015}:
\begin{eqnarray}
\!\!\!i^{(\beta=0)}_{P_G[X]}=\log \mathrm{SNR}- \frac{1}{2} \! \int^{\infty}_{0} \!d
\tau e^{-\tau} \log\left(1+\frac{\tau^2 \tilde{\gamma}^2}{3} \right).
\label{MutInfzero}
\end{eqnarray}
One can check that at small $\tilde{\gamma}$ the expression (\ref{MutInfzero})
reproduces the spectral efficiency $i_{P_G[X]}$, see Eq.~(\ref{SE}), for
$\widetilde{\beta}=0$:
\begin{eqnarray}
i^{(\beta=0)}_{P_G[X]}= \log \mathrm{SNR}-  \frac{\widetilde{\gamma}^2}{3} +{\cal
O}(\widetilde{\gamma}^{\,4}). \label{MutInfResbetazero}
\end{eqnarray}

Let us estimate the spectral efficiency $i_{P[X]}$ for typical fiber optical links
\cite{Essiambre:2010}: $\beta=20\,\mathrm{ps}^2/\mathrm{km}$, $L=1000\,\mathrm{km}$,
$\gamma=1.31(\mathrm{W}\mathrm{km})^{-1}$, $W=100\,\mathrm{GHz}$,
$P_{noise}=QLW/2\pi=5.3\times 10^{-4}\mathrm{mW}$. For these parameters one has
$\widetilde{\beta}=\beta L W^2 \approx  200$, and $g(\widetilde{\beta})\approx
0.42$. Substituting these parameters to Eq.~(\ref{SE}) we obtain
\begin{eqnarray}\label{resultNum}
i_{P[X]}\approx\log \left[\mathrm{SNR}\right]- 7\times
10^{-8}\times\mathrm{SNR}^2\,.
\end{eqnarray}

\begin{figure}[h]
\begin{center}
\includegraphics[width=6cm]{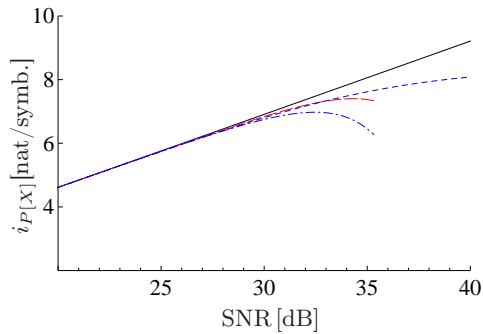}
\begin{picture}(0,0)
\put(-102,-10){\text{$\mathrm{SNR}\,[\mathrm{dB}]$}}
\put(-182,25){\rotatebox{90}{\text{$i_{P[X]}$[nat/symb.]}}}
\end{picture}
\end{center}
\caption{\label{figure2} The spectral efficiency $i_{P[X]}$ for different
$\widetilde{\beta}$. The solid black line, red long-dashed line, blue dashed, blue
dashed-dotted line correspond to $i_{P[X]}$ for a linear channel (Shannon's result),
channel for the dispersion $\widetilde{\beta}=200$, see Eq.~(\ref{resultNum}),
nondispersive channel Eq.~(\ref{MutInfzero}), and the expansion
(\ref{MutInfResbetazero}), respectively. }
\end{figure}

\begin{figure}[h]
\begin{center}
\includegraphics[width=6cm]{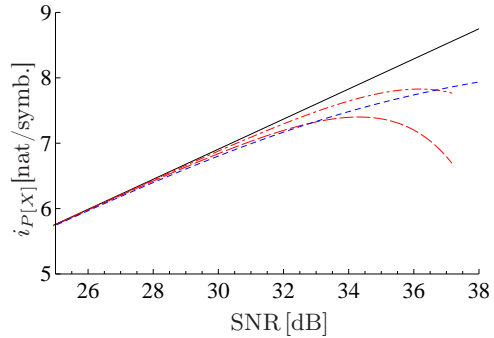}
\begin{picture}(0,0)
\put(-102,-10){\text{$\mathrm{SNR}\,[\mathrm{dB}]$}}
\put(-185,25){\rotatebox{90}{\text{$i_{P[X]}$[nat/symb.]}}}
\end{picture}
\end{center}
\caption{\label{figure3} The spectral efficiency $i_{P[X]}$ for different
$\widetilde{\beta}$. The solid black line, red long-dashed line, blue dashed, red
dashed-dotted line correspond to $i_{P[X]}$ for a linear channel (Shannon's result),
channel for the dispersion $\widetilde{\beta}=200$, see Eq.~(\ref{resultNum}),
nondispersive channel Eq.~(\ref{MutInfzero}), and  the channel for the dispersion
$\widetilde{\beta}=800$, see Eq.~(\ref{resultNum}), respectively. }
\end{figure}

The behavior of  the spectral efficiency for different channels is plotted in
Fig.~{\ref{figure2}}.
The result (\ref{resultNum}) is plotted by the red long-dashed line, the exact
result for the nondispersive channel (\ref{MutInfzero}) and its expansion
(\ref{MutInfResbetazero}) are plotted by the blue dashed  and blue dashed-dotted
lines, respectively. The solid black line corresponds to the Shannon's result
\begin{eqnarray}\label{MIShannon}
i^{SH}_{P[X]}= \log (1+\mathrm{SNR})
\end{eqnarray}
for a linear channel $\tilde{\gamma}=0$. One can see that when $\mathrm{SNR}
\lesssim 300$ ($\mathrm{SNR} \lesssim 25\, \mathrm{dB}$, i.e. $\tilde{\gamma}
\lesssim 0.2$) the spectral efficiency for different channels is close to the
Shannon's result (\ref{MIShannon}). For $\mathrm{SNR}$ large than $25\, \mathrm{dB}$
one observes different a behavior for different regimes. The spectral efficiency
$i^{(\beta=0)}_{P_G[X]}$, see Eq.~(\ref{MutInfzero}), is the nondecreasing function
of the parameter $\mathrm{SNR}$ whereas its expansion (\ref{MutInfResbetazero}) in
$\tilde{\gamma}$ starts decreasing at $\mathrm{SNR} \approx 32\, \mathrm{dB}$. This
decreasing is explained by eliminating of higher terms of expansion in
$\tilde{\gamma}$. It is interesting that the spectral efficiency  for the channel
with dispersion for $\widetilde{\beta}=200$, see Eq.~ (\ref{resultNum}), is greater
than the exact result (\ref{MutInfzero}) for zero dispersion in the region
$\mathrm{SNR} \lesssim 33 \, \mathrm{dB}$, see Fig.~\ref{figure3}.
One can also see that the spectral efficiency  for $\widetilde{\beta}=800$ depicted
by the red dashed-dotted line in Fig.~\ref{figure3} is greater than the exact result
(\ref{MutInfzero}) in the region $\mathrm{SNR} \lesssim 37 \, \mathrm{dB}$.
Increasing parameter $\widetilde{\beta}$ the first nonlinear correction (see
Eq.~(\ref{SE}) and asymptotics (\ref{g-functionlarge}) of the function
$g(\widetilde{\beta})$) goes to zero as $\tilde{\gamma}^2
\log(\widetilde{\beta})/\widetilde{\beta}$. Therefore for larger $\widetilde{\beta}$
the result ~(\ref{SE}) is closer to Shannon's result (\ref{MIShannon}) than the
result (\ref{MutInfzero}) in wider region in $\mathrm{SNR}$.

Let us consider the applicability region of our result (\ref{SE}). To calculate the
spectral efficiency~(\ref{SE}) we have used the perturbative expansion of the
normalization factor $\Lambda_1$ in the parameter $\tilde{\gamma}$. Formally, the
applicability region of our result (\ref{SE}) is defined by the conditions that the
found correction must be much less than the main term ($\log \mathrm{SNR}$ in our
case), and also the next correction of order of $\tilde{\gamma}^4$ must be much less
than the correction of order of $\tilde{\gamma}^2$. We can estimate the next
correction in the dispersive case using the next correction in $\widetilde {\gamma}$
for the nondispersive case. Performing an expansion  in Eq.~(\ref{MutInfzero}) in
$\tilde {\gamma}$ we derive that the next correction has the form $+2 \tilde
{\gamma}^4/3$. However, for the dispersive case instead of $\tilde{\gamma}^2$ we
have an additional suppression factor $g(\widetilde{\beta})\leq 1$. And there is an
indication that at large $\widetilde{\beta}$, the effective parameter of the
perturbative series is $\tilde{\gamma}^2\log(\widetilde{\beta})/{\widetilde{\beta}}$
rather than $\tilde{\gamma}^2$: see Eq.~(\ref{g-functionlarge}). Physically it means
that in the case of large $\widetilde{\beta}$ the dispersion leads to signal
spreading in time domain. It results in the amplitude decreasing and thereby
decreasing of the nonlinear term in the equation (\ref{startingCannelEqt}).
Therefore the effective expansion parameter should be suppressed at large
$\widetilde{\beta}$. And we can estimate the next correction in the dispersive case
as $\sim +(g(\widetilde{\beta})\tilde{\gamma}^2 )^2$. Therefore by increasing the
parameter $\widetilde{\beta}$ we increase the $\mathrm{SNR}$ region of applicability
of our result ~(\ref{SE}). Providing the validity of the indication about the
parameter of the perturbative series our result ~(\ref{resultNum}) for
$\widetilde{\beta}=200$ is applicable in the region $\mathrm{SNR} \lesssim 30 \,
dB$, whereas for $\widetilde{\beta}=800$ the region is $\mathrm{SNR} \lesssim 35 \,
dB$.


\section{ Conclusion}
\label{SectionConclusion}

We have derived the analytical expression for the  mutual information $I_{P[X]}$ of
the channel modelled by the nonlinear Schr\"odinger equation with the additive
Gaussian noise at large $\mathrm{SNR}$. We have calculated analytically the first
nonlinear correction to the mutual information in the nonlinearity parameter
$\tilde{\gamma}=\gamma L P_{ave}$. We have analyzed the obtained result for
different values of the dispersion parameter $\beta$, and we have shown that there
is the region in parameter $\mathrm{SNR}$ where the spectral efficiency (\ref{SE})
for nonzero dispersion channel is greater than the exact result (\ref{MutInfzero})
for the nondispersive channel. We  have also shown that our result for the spectral
efficiency (\ref{SE}) for nonzero dispersion approaches  the expression
(\ref{MutInfResbetazero}) derived in Ref.~\cite{TTKhR:2015} in the case when $\beta$
tends to zero.


\emph{\it Acknowledgments}

The general expressions for the entropies and the mutual information have been
obtained with the support of the Russian Science Foundation (RSF) (grant No.
16-11-10133). Part of the work (perturbative calculation of the mutual information)
was supported by the Russian Foundation for Basic Research (RFBR), Grant No.
16-31-60031/15. A.~V.~Reznichenko thanks the President program for support. The work
of S.~K.~Turitsyn was supported by the grant of the Ministry  of Education and
Science of the Russian Federation (agreement No. 14.B25.31.0003) and the EPSRC
project UNLOC.


\end{document}


\hskip 2cm
\date{\today}
\title{Supplementary Materials for the article: Calculation of mutual information for nonlinear communication channel at large SNR}

\author{I.~S.~Terekhov}
\email[E-mail: ]{I.S.Terekhov@gmail.com} \affiliation{Budker Institute of  Nuclear
Physics of Siberian Branch Russian Academy of Sciences, Novosibirsk, 630090 Russia}
\affiliation{Novosibirsk State University, Novosibirsk, 630090 Russia}
\author{A.~V.~Reznichenko}
\email[E-mail: ]{A.V.Reznichenko@inp.nsk.su}
\affiliation{Budker Institute of  Nuclear Physics of Siberian Branch Russian Academy of Sciences, Novosibirsk, 630090 Russia}
\affiliation{Novosibirsk State University, Novosibirsk, 630090 Russia}
\author{S.~K.~Turitsyn}
\email[E-mail: ]{s.k.turitsyn@aston.ac.uk} \affiliation{Aston Institute of Photonics
Technologies, Aston University, Aston Triangle, Birmingham, B4 7ET, UK}
\affiliation{Novosibirsk State University, Novosibirsk, 630090 Russia}


\maketitle

\section{Classical solution of the Euler-Lagrange equation $\delta S [\Psi]=0$}

In Ref.\cite{Terekhov:2014} the following representation for the
conditional probability was obtained:
\begin{widetext}
\begin{eqnarray}
P[Y|X]=\Lambda \exp\left\{-\frac{S[\Psi]}{Q}\right\},\qquad
\Lambda=\int^{\phi(z=L)=0}_{\phi(z=0)=0} {\cal D} \phi \exp\left\{
-\frac{1}{Q}\left(S[\Psi +\phi]-S[\Psi]\right)\right\}\,,\label{SMPYXquasiclassic}
\end{eqnarray}
\end{widetext}
where the action $S[\psi]$ in (\ref{SMPYXquasiclassic}) reads in the time domain
\begin{eqnarray}\label{SMactionint}
&&S[\psi]=\int_0^L dz\int_{T}dt\left|{\cal L}[\psi(z,t)]\right|^2\,,\qquad {\cal
L}[\psi]=\partial_z \psi(z,t)+i\beta\partial^2_{t}\psi(z,t)-i\gamma \psi(z,t)
|\psi(z,t)|^2\,.
\end{eqnarray}
Here $T$ is a time interval containing both signals $X(t)$ and $Y(t)$. In the
following discretization scheme we will use the relations between $T$ and
discretization intervals in the time domain ($\delta_t$ for the ``dense'' time grid
with $M'$ intervals and $\tilde{\delta}_t$ for the ``coarse'' time sub-grid with $M$
intervals, see Ref.~\cite{TheArticle:2016}) and in the frequency domain ($\delta_{\omega}$):
\begin{eqnarray}\label{Trelations}
&&T=1/\delta_{\omega}=M' \delta_t = M \tilde{\delta}_t=2\pi M/W =2 \pi M'/W'.
\end{eqnarray}

The function $\Psi$ in Eq.~(\ref{SMPYXquasiclassic}) is the solution (referred to as
``the classical solution'') of the Euler-Lagrange equation $\delta S [\Psi]=0$ with
the boundary conditions: $\Psi (0)=X\,,\quad \Psi (L)=Y$. In the time domain this
equation for $\Psi (z,t)$ has a notedly simple form
\begin{eqnarray}\label{classicalTrajectorytime}
&&\left(\partial_z+i\beta\partial^2_t -2 i \gamma |\Psi(z,t)|^2\right){\cal
L}[\Psi(z,t)]+i\gamma \Psi^2(z,t) \overline{{\cal L}[\Psi(z,t)]}=0, \nonumber \\&&
{\cal L}[\Psi(z,t)]=\left(\partial_z+i\beta\partial^2_t - i \gamma
|\Psi(z,t)|^2\right)\Psi(z,t),
\end{eqnarray}
and the function $\Psi(z,t)$ obeys the boundary conditions: $\Psi(0,t)=X(t)$,
$\Psi(L,t)=Y(t)$. The bar in Eq.~(\ref{classicalTrajectorytime}) and hereafter means
complex conjugation.

It is convenient to introduce the function $\Phi(z,t)$ which is the solution of the
nonlinear Schr\"odinger equation (NLSE)
\begin{eqnarray}\label{SMstartingCannelEq}
&&\partial_z \Phi(z,t)+i\beta\partial^2_{t}\Phi(z,t)-i\gamma\, \Phi(z,t)
|\Phi(z,t)|^2=0 \,
\end{eqnarray}
with the boundary condition $\Phi(0,t)=X(t)$. One can see that the equation
(\ref{SMstartingCannelEq}) can be written as
\begin{eqnarray}\label{Phiproblem}
&&{\cal L}[\Phi(z,t)]=0.
\end{eqnarray}
Therefore it is obvious that the function $\Phi(z,t)$ obeys the equation
Eq.~(\ref{classicalTrajectorytime}) and the boundary condition at $z=0$, but it does
not obey the boundary condition at $z=L$. It globally minimizes the action as well:
$S[\Phi(z,t)]=0$. Since we imply that the noise power is much less than the signal
power we can present the solution of Eq. (\ref{classicalTrajectorytime}) in the form
\begin{eqnarray}\label{PsiToKappa}
\Psi(z,t)=\Phi(z,t)+\varkappa(z,t)\,,
\end{eqnarray}
where the function $\varkappa(z,t)$ is  of order of $\sqrt{Q}$ for  unsuppressed
configurations $\Psi(z,t)$. Therefore we substitute  the function $\Psi$ in the form
(\ref{PsiToKappa}) to the Eq. (\ref{classicalTrajectorytime}), then linearizing
Eq.~(\ref{classicalTrajectorytime}) in $\varkappa(z,t)$ we obtain the following
linear problem on $\varkappa(z,t)$:
\begin{eqnarray}\label{kappaequation}
&&\left(\partial_z+i\beta\partial^2_t -2 i \gamma |\Phi(z,t)|^2\right)
l[\varkappa(z,t)]+i\gamma \Phi^2(z,t) \overline{l[\varkappa(z,t)]}=0, \nonumber \\&&
l[\varkappa(z,t)]=\left(\partial_z+i\beta\partial^2_t\right)\varkappa(z,t)- i \gamma
\left(2 \varkappa(z,t)|\Phi(z,t)|^2+ \overline{\varkappa(z,t)}\Phi^2(z,t) \right),
\end{eqnarray}
with the boundary conditions
\begin{eqnarray}\label{kappacondition}
&&\varkappa(z=0,t)=0,\qquad  \varkappa(z=L,t)= Y(t)- \Phi(L,t) \equiv \delta Y(t).
\end{eqnarray}
Being rewritten explicitly the Eq.~(\ref{kappaequation}) has the following form
\begin{eqnarray}\label{SMkappaequationfull}
&&\left[\left(\partial_z+i\beta\partial^2_t\right)^2- 4 i \gamma|\Phi|^2
\left(\partial_z+i\beta\partial^2_t\right)-5\gamma^2  |\Phi|^4+4\beta \gamma
\left(\Phi \frac{\partial^2 \bar{\Phi}}{\partial t^2}+\left|\frac{\partial
\Phi}{\partial t}\right|^2+\frac{\partial |\Phi|^2}{\partial t}\,
\partial_t \right) \right]\varkappa(z,t)+\nonumber \\
&& 2\gamma \left[-\gamma \Phi^2 |\Phi|^2+\beta \Phi^2\,
\partial^2_{t}+\beta\frac{\partial \Phi^2}{\partial t}\,\partial_{t}+
\beta \left( \frac{\partial \Phi}{\partial t} \right)^2
\right]\overline{\varkappa(z,t)}=0,
\end{eqnarray}
where we have used that $\Phi(z,t)$ is the solution of Eq.
(\ref{SMstartingCannelEq}).

In the frequency domain the Euler-Lagrange equation $\delta S [\Psi]=0$ for the
classical solution $\Psi_{\omega}(z)$ has the form \cite{Terekhov:2014}:
\begin{eqnarray}\label{classicalTrajectoryomega}
&& \left(\partial_z-i \beta \omega^2 \right)^2 \Psi_{\omega}(z)- i \gamma \int
\frac{d \omega_1 d \omega_2 d
\omega_3}{(2\pi)^2}\delta\left(\omega+\omega_3-\omega_1-\omega_2\right)\Big\{ 4
\Psi_{\omega_2}(z) \bar{\Psi}_{\omega_3}(z)\Big(\partial_z-i \beta
\omega^2_1\Big)\Psi_{\omega_1}(z)- \nonumber \\&& i \beta
\left(\omega^2-\omega^2_1-\omega^2_2+\omega^2_3\right)
\Psi_{\omega_1}(z)\Psi_{\omega_2}(z)\bar{\Psi}_{\omega_3}(z) \Big\}- \nonumber
\\&&3\gamma^2 \int \frac{d \omega_1 d \omega_2 d \omega_4 d \omega_5 d
\omega_6}{(2\pi)^4}\delta\left(\omega_1+\omega_2+\omega_4-\omega_5-\omega_6-\omega\right)
\Psi_{\omega_1}(z)\Psi_{\omega_2}(z)\Psi_{\omega_4}(z)\bar{\Psi}_{\omega_5}(z)
\bar{\Psi}_{\omega_6}(z)=0,
\end{eqnarray}
with the boundary conditions: $\Psi_{\omega}(0)=X(\omega)$,
$\Psi_{\omega}(L)=Y(\omega)$.
The equation (\ref{SMkappaequationfull}) for the function $\varkappa$ can be
rewritten in the frequency domain as follows:
\begin{eqnarray}\label{SMkappaequationomega}
&& \left(\partial_z-i \beta \omega^2 \right)^2 \varkappa_{\omega}(z)- i \gamma \int
\frac{d \omega_1 d \omega_2 d
\omega_3}{(2\pi)^2}\delta\left(\omega+\omega_3-\omega_1-\omega_2\right)\Big\{ 4
\Phi_{\omega_2}(z) \bar{\Phi}_{\omega_3}(z)\Big(\partial_z-i \beta
\omega^2_1\Big)\varkappa_{\omega_1}(z)- \nonumber \\&& i \beta
\left(\omega^2-\omega^2_1-\omega^2_2+\omega^2_3\right)\Phi_{\omega_2}(z)\left[2
\varkappa_{\omega_1}(z)\bar{\Phi}_{\omega_3}(z)+\bar{\varkappa}_{\omega_3}(z)\Phi_{\omega_1}(z)
\right] \Big\} - \nonumber
\\&&\gamma^2 \int \frac{d \omega_1 d \omega_2 d \omega_4 d \omega_5 d
\omega_6}{(2\pi)^4}\delta\left(\omega_1+\omega_2+\omega_4-\omega_5-\omega_6-\omega\right)
\Phi_{\omega_2}(z)\Phi_{\omega_4}(z)\bar{\Phi}_{\omega_5}(z)\times \nonumber \\&&
\times \left[5\varkappa_{\omega_1}(z)
\bar{\Phi}_{\omega_6}(z)+2\bar{\varkappa}_{\omega_6}(z) \Phi_{\omega_1}(z) \right]
=0.
\end{eqnarray}
Of course, Eq.~ (\ref{SMkappaequationomega})  can be obtained from
Eq.~(\ref{classicalTrajectoryomega}) by the linearization procedure using
Eq.~(\ref{SMstartingCannelEq}) for the function $\Phi_{\omega}(z)$.

{In our model there are two grids both in frequency and time domains. In the
frequency domain $W$ we have $M$ grid points corresponding to the meaning complex
channels and $M'-M$ points in the domain $W' \setminus W$ corresponding to the
channels with zero initial signal $X(\omega)$. These grids in the frequency domain
relate with the coarse and dense grids in the time domain. Grid points in the coarse
time grid are separated by intervals $\tilde{\delta}_{t}=T/M=2\pi/W$ and carry
information about $M$ meaning channels. The dense time grid contains $M'$ points
separated by intervals ${\delta}_{t}=T/M'=2\pi/W'$, see Eq.~(\ref{Trelations}). The
signals in $M'-M$ remnant points are uniquely defined by the signals on the coarse
grid. Therefore} to obtain Eqs.~(\ref{classicalTrajectoryomega}) and
(\ref{SMkappaequationomega}) in our frequency discretization scheme  we should
perform the following substitutions:
\begin{eqnarray}\label{SMconttodisc}
&& \partial^2_{t}\Phi(z,t) \rightarrow -\Omega^2_{k}\Phi_{\omega_k}(z), \qquad
\delta\left(\omega_{i_1}+\omega_{i_2}+\ldots\right) \rightarrow
\frac{\Delta_{(M')}\left(i_1+i_2+\ldots \right)}{2\pi \delta_{\omega}},\qquad \int
\frac{d\omega}{2\pi}\ldots \rightarrow \delta_{\omega} \sum_{j'=0}^{M'-1},
\end{eqnarray}
\begin{eqnarray} \label{SMOmega}
&& \qquad \omega_{n'}=-W'/2+ 2\pi \delta_{\omega}n', \quad \Omega_{n'}=2\sin[\pi n'/
M']{M'}/{T}=2 M'\delta_{\omega} \sin[\pi n'/ M']\,, \quad n'=0,1,\ldots,M'-1,
\end{eqnarray}
\begin{eqnarray}\label{SMdiscretedelta}
&&\Delta_{(M)}(k)=\frac{1}{M}\sum^{M-1}_{n=0}\exp\left\{-2\pi i \frac{n}{M}k\right\}=\sum^{\infty}_{m=-\infty} \delta_{k, m M}.
\end{eqnarray}

\section{Factorization of pre-exponential factor $\Lambda$}
Let us consider the separation of different scales in the conditional PDF $P[Y|X]$
when PDF $P[Y|X]$ is considered under an integral over $X$ together with the initial
signal PDF $P[X]$ which has the following form in the frequency domain:
\begin{eqnarray}
&\mspace{-20mu}P[\vec{X}]=  P^{(M)}_{X}[\vec{X}_1]
\delta\left(\vec{X}_2\right),\,\label{SMPX2}
\end{eqnarray}
where $\delta\left(\vec{X}_2\right)$ means $2(M'-M)$-dimensional delta-function
corresponding $M'-M$ complex remnant channels. We remind that the vector notations
(introduced in the main text of our manuscript, see \cite{TheArticle:2016}, Section
III) ${\vec X}={\vec X}_1 \oplus {\vec X}_2$ relate to the frequency domain: ${\vec
X}_1$ is $2M$-dimensional vector corresponding to $M$ meaning complex channels in
the frequency domain $W$, whereas ${\vec X}_2$ is $2(M'-M)$-dimensional vector
corresponding to remnant  $M'-M$ complex channels in the frequency domain
$W'\setminus W$. Now we are going to demonstrate the factorization property of
$P[Y|X]$ in the time domain.

In the leading order in $1/\mathrm{SNR}$ the conditional PDF $P[Y|X]$ can be
obtained by substituting $\Psi$ in the form $\Psi=\Phi+\varkappa$ to the
Eq.~(\ref{SMPYXquasiclassic}). After obvious transformations (see Section III in the
main text of our manuscript \cite{TheArticle:2016}) we obtain:
\begin{eqnarray}\label{PYXsimpl}
&&P[Y|X] \approx \Lambda[X]\, e^{-{S}_{2}[\varkappa]/Q}\,,
\end{eqnarray}
where the action ${S}_2[\phi]$ is quadratic functional in $\phi$ and it reads (see
Eq.~(23) in \cite{TheArticle:2016})
\begin{eqnarray}\label{actionsimpl}
&&{S}_2[\phi]= {\delta_{t}}\!\!\sum^{M'-1}_{k=0} \Delta \sum^{N-1}_{n=1} \,{\cal
L}_{eff}[\phi(z_n,t_k)],
\end{eqnarray}
where ${\delta_{t}}=T/M'$ is the discretization parameter in the time domain: $t_k=k
\delta_{t}$, $k=0,1,\ldots, M'-1$. In Eq.~(\ref{actionsimpl}) $\Delta=L/N$ is the
distance discretization parameter: $z_n=n \Delta$, $n=1,2,\ldots, N-1$. In the
action (\ref{actionsimpl}) we have introduced the ``Lagrangian'':
\begin{eqnarray}\label{actionsimp2}
&& {\cal L}_{eff}[\varkappa]=\left|\partial_z
\varkappa(z_n,t_{k})+i\beta\partial^2_{t}\varkappa(z_n,t_{k})-i\gamma\, \Big( 2
\varkappa(z_n,t_{k}) |\Phi(z_n,t_{k})|^2 +
\bar{\varkappa}(z_n,t_{k})\Phi^2(z_n,t_{k})\Big)\right|^2.
\end{eqnarray}
Here derivatives should be regarded as difference derivatives in our
discretization scheme. The function $\varkappa(z,t)$ in the exponent
(\ref{PYXsimpl}) is the solution of the Euler-Lagrange equation $-\delta {\cal
L}_{eff}[\varkappa]/{\delta \bar{\varkappa}}=0$ which coincides with
Eq.~(\ref{SMkappaequationfull}) with the boundary conditions $\varkappa(z=0,t)=0$,
$\varkappa(z=L,t)=Y(t)-\Phi(L,t)$. Here $\Phi(L,t)=\hat{\mathrm{L}} X (t)$ is the
solution of NLSE with the zero noise and with the input boundary condition, see
Eq.~(\ref{Phiproblem}). The normalization factor $\Lambda[X]$ has the form
\begin{eqnarray}\label{PathIntegrallambda}
&&\Lambda[X]=\int\limits_{\phi (0,t)=0}^{\phi (L,t)=0}\!\!\!\!\!{\cal D}\phi\,
\exp\left\{-\frac{\delta_{t}}{Q}\!\!\sum^{M'-1}_{k=0}\Delta \sum^{N-1}_{n=1}  {\cal
L}_{eff}[\phi(z_n,t_{k})] \right\}\,.
\end{eqnarray}
Note that the sum in Eq.~(\ref{actionsimpl}) is performed over the dense time grid.
To demonstrate the factorization we have to separate the scales in the action into
the coarse and dense parts. In other words, we have to separate the summation over
$M$ meaning channels and $M'-M$ remnant channels. The scale separation procedure in
some sense is similar to Wilson's renormalization procedure for the Lagrangian
${\cal L}_{eff}[\varkappa]$, see \cite{Peskin:1995}. But in our approximation the
Lagrangian (\ref{actionsimp2}) is quadratic functional in $\varkappa$ that is why
there are no corrections to the effective action when we perform integration over
remnant $2(M'-M)$ degrees of freedom $\varkappa(z,t_{k})$ where $t_{k}$  runs
through values only on the dense grid without the coarse sub-grid. Let us
demonstrate this fact.

First we perform the separation of variables:
\begin{eqnarray}\label{separation}
&&\varkappa(z,t_{k})=\varkappa^{(c)}(z,t_{k})+\varkappa^{(d)}(z,t_{k}),
\\&&
Y(t_{k})=Y^{(c)}(t_{k})+Y^{(d)}(t_{k}) \nonumber,
\end{eqnarray}
where $\varkappa^{(c)}(z,t_{k})$, or $Y^{(c)}(t_{k})$, is completely defined only by
the values $\varkappa^{(c)}(z,T_{\tilde{i}})$, or $Y^{(c)}(T_{\tilde{i}})$, on the
coarse time grid $T_{\tilde{i}}= \tilde{i} \tilde{\delta}_{t}$,
$\tilde{i}=0,1,\ldots, M-1$. Here and below superscript ``(c)'' means ``coarse''
variable. In other words, the function $\varkappa^{(c)}(z,t_{k})$ evaluated in all
grid points is the interpolation of some order (i.e. the interpolating polynomial
degree) $N_0>2$  calculated on the base of values $\varkappa^{(c)}(z,T_{\tilde{i}})$
of the coarse time grid. The function $\varkappa^{(c)}(z,t_{k})$ coincides with
$\varkappa(z,T_{\tilde{i}})$ when $t_{k}$ falls on the coarse time grid
$T_{\tilde{i}}$ (i.e. $k= [\tilde{i} M'/M]$, $\tilde{i}=0,1,\ldots, M-1$), i.e.
$\varkappa^{(d)}(z,t_{k})=0$ on the coarse grid. In other grid points of the dense
grid the function $\varkappa^{(c)}(z,t_{k})$ smoothly interpolates the values of
$\varkappa(z,t_{k})$ with interpolation order $N_0 > 2$: $\varkappa^{(d)} = {\cal
O}(\tilde{\delta}^{N_0}_t)$ and $\partial^2_{t}\varkappa^{(d)}(z,t_{k})= {\cal
O}(\tilde{\delta}^{N_0-2}_t)$, where we have used that
$\partial^2_{t}\varkappa(z,t_{k})=\partial^2_{t}\varkappa^{(c)}(z,t_{k})
+\partial^2_{t}\varkappa^{(d)}(z,t_{k})$, here the derivatives are assumed as the
difference derivatives on the dense grid. The boundary conditions are as follows:
\begin{eqnarray}\label{separationbc}
&&\varkappa^{(c)}(0,t_{k})=0,\qquad
\varkappa^{(c)}(L,t_{k})=Y^{(c)}(t_{k})-\Phi^{(c)}(L,t_{k}), \nonumber
\\&& \varkappa^{(d)}(0,t_{k})=0, \qquad \varkappa^{(d)}(L,t_{k})=Y^{(d)}(t_{k})-\Phi^{(d)}(L,t_{k}),
\end{eqnarray}
where we have used that $\Phi(L,t_{k})$ has the ``coarse'' and ``dense'' parts as
well:
\begin{eqnarray}
&& \Phi(z,t_{k})=\Phi^{(c)}(z,t_{k})+\Phi^{(d)}(z,t_{k}), \quad
\Phi^{(d)}(z,t_{k})={\cal O}(\tilde{\delta}^{N_0}_t).
\end{eqnarray}
Note that if we consider (\ref{PYXsimpl}) under the integral over ${\cal D} X$ with
the initial signal PDF (\ref{SMPX2}), then the function $\Phi(z,t_{k})$ is the
(nonlinear) function of the input signal only on the coarse time grid
$X(T_{\tilde{i}})$. This means that the ``dense'' part $\Phi^{(d)}(z,t_{k})={\cal
O}(\tilde{\delta}^{N_0}_t)$ is always small for sufficiently large $M$.

Now we insert the representation (\ref{separation}) in the action
(\ref{actionsimpl}). The action fractionizes into three parts
\begin{eqnarray}\label{Skappa}
&& {S}_2[\varkappa]={\delta_{t}}\!\!\sum^{M'-1}_{k=0}\!\Delta \sum^{N-1}_{n=1} {\cal
L}_{eff}[\varkappa(z_n,t_{k})]={\delta_{t}}\!\!\sum^{M'-1}_{k=0}\!\Delta
\sum^{N-1}_{n=1}   {\cal L}_{eff}[\varkappa^{(c)}(z_n,t_{k})]+
{\delta_{t}}\!\!\sum^{M'-1}_{k=0}\!\Delta \sum^{N-1}_{n=1}  {\cal
L}_{eff}[\varkappa^{(d)}(z_n,t_{k})]+\nonumber \\&&
{\delta_{t}}\!\!\sum^{M'-1}_{k=0}\!\Delta \sum^{N-1}_{n=1} {\cal
L}_{int}[\varkappa^{(c)}(z_n,t_{k}),\varkappa^{(d)}(z_n,t_{k})],
\end{eqnarray}
where the third part with interaction of ``coarse'' ($\varkappa^{(c)}$) and
``dense'' ($\varkappa^{(d)}$) degrees of freedom contains  Lagrangian
\begin{eqnarray} \label{Linteraction}
&&{\cal L}_{int}[\varkappa^{(c)}(z,t_{k}),\varkappa^{(d)}(z,t_{k})]=\nonumber
\\&&\left(\partial_z
\varkappa^{(c)}(z,t_{k})+i\beta\partial^2_{t}\varkappa^{(c)}(z,t_{k})-i\gamma\,
\Big( 2 \varkappa^{(c)}(z,t_{k}) |\Phi(z,t_{k})|^2 +
\bar{\varkappa}^{(c)}(z,t_{k})\Phi^2(z,t_{k})\Big)\right)\times \nonumber \\&&
\left(\partial_z
\bar{\varkappa}^{(d)}(z,t_{k})-i\beta\partial^2_{t}\bar{\varkappa}^{(d)}(z,t_{k})+i\gamma\,
\Big( 2 \bar{\varkappa}^{(d)}(z,t_{k}) |\Phi(z,t_{k})|^2 +
{\varkappa}^{(d)}(z,t_{k})\bar{\Phi}^2(z,t_{k})\Big)\right)+c.c.
\end{eqnarray}
Here ``$c.c.$''  means the same complex conjugated term.

The first part in the r.h.s. of Eq.~(\ref{Skappa}) can be simplified as follows:
\begin{eqnarray}\label{firstterm}
&& {\delta_{t}}\!\!\sum^{M'-1}_{k=0}\!\Delta \sum^{N-1}_{n=1}  {\cal
L}_{eff}[\varkappa^{(c)}(z_n,t_{k})]={\tilde{\delta}_{t}}\sum^{M-1}_{\tilde{i}=0}\!\Delta
\sum^{N-1}_{n=1} {\cal L}_{eff}[\phi(z_n,T_{\tilde{i}})]\left(1+{\cal O}
(\tilde{\delta}^{N_0-2}_{t})\right),
\end{eqnarray}
where $\tilde{\delta}_t=\frac{2\pi}{W}= \delta_t \Big[M'/M\Big]$ is the grid spacing
of the coarse time grid. Here we have replaced every term under the sum over the
dense grid with its average value on the coarse grid. The accuracy in
Eq.~(\ref{firstterm}) is governed by interpolation order of the second derivative in
time.

The third part in the r.h.s. of Eq.~(\ref{Skappa}) can be integrated (summed) over
$z$ by part resulting in the following expression:
\begin{eqnarray} \label{Sint}
&& {\delta_{t}}\!\!\sum^{M'-1}_{k=0}\!\Delta \sum^{N-1}_{n=1}   {\cal
L}_{int}[\varkappa^{(c)}(z_n,t_{k}),\varkappa^{(d)}(z_n,t_{k})]=
{\delta_{t}}\!\!\sum^{M'-1}_{k=0}\!\Delta \sum^{N-1}_{n=1}
\left(\bar{\varkappa}^{(d)}(z_n,t_{k})\frac{\delta {\cal
L}_{eff}[\varkappa^{(c)}]}{\delta \varkappa}+h.c. \right) + S_{surf},
\end{eqnarray}
where the variation $-\delta {\cal L}_{eff}[\varkappa^{(c)}]/{\delta \varkappa}$ is
linear in $\varkappa^{(c)}$, and it represents the l.h.s. of the  Euler-Lagrange
equation (\ref{SMkappaequationfull}), i.e. for the function $\varkappa^{(c)}$ we
obtain $\delta {\cal L}_{eff}[\varkappa^{(c)}]/{\delta \varkappa}= {\cal
O}(\tilde{\delta}^{N_0-2}_{t})$, i.e. it is always small. The term $S_{surf}$
results from the surface term in integration by part over $z$ in Eq.~(\ref{Sint}),
and taking into account the boundary conditions (\ref{separationbc}) it reads
\begin{eqnarray} \label{Ssurf}
&&S_{surf}={\delta_{t}}\sum^{M'-1}_{k=0}\left[
\bar{Y}^{(d)}(t_{k})-\bar{\Phi}^{(d)}(L,t_{k})\right]\Big(\partial_z
\varkappa^{(c)}(L,t_{k})+i\beta\partial^2_{t}\varkappa^{(c)}(L,t_{k})- \nonumber
\\&& i\gamma\, \Big( 2 \varkappa^{(c)}(L,t_{k}) |\Phi^{}(L,t_{k})|^2 +
\bar{\varkappa}^{(c)}(L,t_{k})\Phi^{2}(L,t_{k})\Big)\Big)+ c.c.
\end{eqnarray}
We can omit the surface term (\ref{Ssurf}), since  it is linear both in the
``coarse'' and ``dense'' variables, but they are orthogonal when integrating over
$t$ (they have not intersecting supports in the frequency domain). It is obvious for
the first two terms in the parentheses in Eq.~(\ref{Ssurf}). The last terms
containing $\Phi^{}(L,t_{k})$ are ``coarse'' variables as well: we can replace
$\Phi^{}(L,t_{k})$ with $\Phi^{(c)}(L,t_{k})$ with the interpolation accuracy ${\cal
O}(\tilde{\delta}^{N_0}_{t})$ and then replace $\Phi^{(c)}(L,t_{k})$ with
$Y^{(c)}(t_k)$ with the accuracy ${\cal O}(\sqrt{Q})$ (we remind that $\varkappa$ in
Eq.~(\ref{separationbc}) is of order of $\sqrt{Q}$). Then we can replace
$Y^{(c)}(t_k)$ with the constants inside the whole interval of an coarse space with
the interpolation accuracy ${\cal O}(\tilde{\delta}_{t})$ and now  use the
orthogonality of the ``coarse'' and ``dense'' variables.

To summarize, with the accuracy of our interpolation ${\cal
O}(\tilde{\delta}_t)={\cal O}(1/M)$ we can omit the interaction term (\ref{Sint}),
and our action fractionizes into ``coarse'' and ``dense'' parts:
\begin{eqnarray} \label{Ssum}
&& {S}_2[\varkappa]\approx {S}_2[\varkappa^{(c)}]+{S}_2[\varkappa^{(d)}],
\end{eqnarray}
where we have separated ``coarse'' and ``dense'' degrees of freedom:
$S_2[\varkappa^{(c)}]$ depends on $Y^{(c)}(t_j)$, and $S_2[\varkappa^{(d)}]$ depends
on $Y^{(d)}(t_j)$ only. Both actions are expressed through the same Lagrangian
(\ref{actionsimp2}) and are represented as the quadratic forms. The coefficients of
these quadratic forms depend on input signal ${X}$ only.

The factorization  of $\Lambda$, see Eq. (38) of Ref.~\cite{TheArticle:2016},  can be shown using the normalization condition:\begin{eqnarray} \label{normalizationPYX1}
1=\int {\cal D} Y P[Y|X]= \Lambda \int {\cal D} Y
\exp\left\{-{S}_2[\varkappa]/Q\right\},
\end{eqnarray}
where we have used that $\Lambda$ does not depend on $Y$ in the leading order in
$1/\mathrm{SNR}$. Taking into account (\ref{normalizationPYX1}) and (\ref{Ssum}) we
obtain
\begin{eqnarray} \label{normalizationPYX2}
\Lambda^{-1}= \int {\cal D} Y \exp\left\{-\frac{{S}_2[\varkappa^{(c)}]}{Q}-
\frac{{S}_2[\varkappa^{(d)}]}{Q}\right\}=\int {\cal D} Y^{(c)}
\exp\left\{-\frac{{S}_2[\varkappa^{(c)}]}{Q}\right\} {\cal D} Y^{(d)} \exp\left\{-
\frac{{S}_2[\varkappa^{(d)}]}{Q}\right\}= \Lambda^{-1}_1 \times \Lambda^{-1}_2,
\end{eqnarray}
or
\begin{eqnarray}\label{SMlambdafactorization}
\Lambda=\Lambda_1 \times \Lambda_2\,.
\end{eqnarray}
Here the normalization factor $\Lambda_1$ depends on the input signal $X$ on the coarse
grid only and reads
\begin{eqnarray}\label{PathIntegral3}
&&\Lambda_1=\!\!\!\!\!\!\!\!\!\int\limits_{\phi (0,t)=0}^{\phi
(L,t)=0}\!\!\!\!\!\Big[{\cal D}\phi(z,t) \Big]_{M}\,
\exp\left\{-\frac{\tilde{\delta}_{t}}{Q}\!\!\sum^{M-1}_{\tilde{i}=0}\!\Delta
\sum^{N-1}_{n=1} {\cal L}_{eff}[\phi(z_n,T_{\tilde{i}})] \right\}\,,\nonumber \\&&
{\cal L}_{eff}[\phi(z_n,T_{\tilde{i}})]=\left|\partial_z \phi(z_n,T_{\tilde{i}})+
i\beta\partial^2_{t}\phi(z_n,T_{\tilde{i}})-i\gamma\, \Big( 2
\phi(z_n,T_{\tilde{i}}) |\Phi(z_n,T_{\tilde{i}})|^2 +
\bar{\phi}(z_n,T_{\tilde{i}})\Phi^2(z_n,T_{\tilde{i}})\Big)\right|^2.
\end{eqnarray}
The measure $\Big[{\cal D}\phi(z,t) \Big]_{M}$ is defined as
\begin{eqnarray} \label{measureintM}
&& \Big[{\cal D} {\phi (z,t)}\Big]_{M}=\lim_{\tilde{\delta}_{t} \rightarrow
0}\lim_{\Delta \rightarrow 0} \Big( \frac{\tilde{\delta}_{t}}{\Delta \pi
Q}\Big)^{M}\prod^{M-1}_{\tilde{j}=0}\prod^{N-1}_{i=1}\Big\{
\frac{\tilde{\delta}_{t}}{\Delta \pi Q}\,\,
dRe\phi(z_i,T_{\tilde{j}})\,dIm\phi(z_i,T_{\tilde{j}})\Big\}.
\end{eqnarray}
In Eq.~(\ref{SMlambdafactorization}) the normalization factor $\Lambda_1$ corresponds
to $M$ meaning complex channels, and $\Lambda_2$ corresponds to $M'-M$ complex
remnant channels. In this demonstration we have used that the quantity $P[Y|X]$ is
considered under the integral over ${\cal D} X$ with the initial signal PDF $P[X]$,
see Eq.~(\ref{SMPX2}). The  accuracy of our factorization is at least ${\cal
O}(\tilde{\delta}_t)={\cal O}(1/M)$.

For illustration of the factorization (\ref{SMlambdafactorization}) let us consider
the factorization property for the conditional PDF $P[Y|X]$ and for $\Lambda$ in two
cases: a linear channel with nonzero dispersion and a nonlinear nondispersive
channel \cite{TTKhR:2015}.

The linear channel ($\gamma=0$) with dispersion  has this exact property notedly
simple in the frequency domain, see e.g. \cite{Terekhov:2014}:
\begin{eqnarray} \label{PYXfactlinear}
&& P[Y|X]=\left(\frac{\delta_{\omega}}{\pi
QL}\right)^{M'}\exp\left\{-\frac{\delta_{\omega}}{QL}
\sum^{M'-1}_{k'=0}\left|Y(\omega_{k'})e^{-i \beta L
\Omega^2_{k'}}-X(\omega_{k'})\right|^2 \right\}= P^{(M)}[Y|X] \times P^{(M'-M)}[Y|X].
\end{eqnarray}
In the linear  case the same factorization is valid for the normalization factor
acquiring the trivial form: $\Lambda=\left({\delta_{\omega}}/{(\pi
QL)}\right)^{M'}=\Lambda_1\times \Lambda_2$, $\Lambda_1=\left({\delta_{\omega}}/{(\pi
QL)}\right)^{M}$, $\Lambda_2=\left({\delta_{\omega}}/{(\pi
QL)}\right)^{M'-M}$.

For the nonlinear ($\gamma$ is arbitrary) nondispersive ($\beta=0$) channel in the
leading order in $1/\mathrm{SNR}$ the factorization property was derived in
\cite{TTKhR:2015}. In the time domain $P[Y|X]$ as well as $\Lambda$ are factorized:
\begin{eqnarray} \label{PYXnondisp}
&& P[Y|X]=\prod^{M'-1}_{j=0} \dfrac{\delta_t} {\pi Q L }
\frac{1}{\sqrt{1+\mu_{j}^2/3}}\exp\left\{- \frac{\delta_t}{QL}
\dfrac{(1+4\mu_{j}^2/3)x^2_j-2\mu_{j} x_j
y_j+y^2_j}{(1+\mu_{j}^2/3)}\right\}\,,
\end{eqnarray}
where $\mu_{j}=\gamma L \left|X(t_j)\right|^2$, and $x_j+iy_j =Y(t_j) e^{-i
\phi^{(X(t_j))}-i\mu_j}-\left|X(t_j)\right|$, with $\phi^{(X(t_j))}$ being the phase of the input signal $X(t_j)$. One can see that in the case the conditional probability is the product of the conditional probabilities of $M'$ intdependent channels in the time domain, therefore the factorization is obvious.

\section{Perturbative calculation of mutual information}

In what follows we will calculate the normalization factor $\Lambda_1$ in the
nonlinear dispersive case in the perturbative expansion in nonlinearity
dimensionless parameter $\widetilde{\gamma}= P_{ave} \gamma L $. We start from the
general expression for the mutual information in the leading order in
$1/\mathrm{SNR}$ obtained in the main text of the manuscript, see
\cite{TheArticle:2016},  Section III:
\begin{eqnarray} \label{SMMutualinformation}
I_{P[X]}= H[X]-M+\int d\vec{X}_1  P^{(M)}_{X}[\vec{X}_1 ]\,\log \Lambda_1[\vec{X}_1
],
\end{eqnarray}
where $\vec{X}_1$ is $2M$ dimensional vector corresponding $M$ complex meaning
channels in the frequency domain. It is worth noting that this representation is
valid for arbitrary nonlinearity but in the leading order in $1/\mathrm{SNR}$. Now
we calculate $\Lambda_1[\vec{X}_1]$ within the perturbation theory in dimensionless
parameter $\tilde{\gamma}$ and perform the averaging over Gaussian input signal PDF
$P^{(M)}_{X}$
\begin{eqnarray}
P^{(M)}_{X}[\vec{X}_1]=P_{G}[\vec{X}_1]=\Lambda_{P} \, e^{- |\vec{X}_1|^2
\delta_{\omega}/P},\, \label{SMPXGaussian}
\end{eqnarray}
where the factor $\Lambda_{P}=({\delta_{\omega}}/{(\pi P)})^{M}$, is consistent with
the normalization condition $\int d \vec{X}_1 P^{(M)}_{X}[\vec{X}_1] =1$. Here
$\delta_{\omega}=W/(2\pi M)=W'/(2\pi M')$ is the grid spacing in the frequency
domain, see Eq.~(\ref{Trelations}).

For short we introduce the notation of averaging over $\vec{X}_1$ as:
\begin{eqnarray}\label{AverNotqtion}
\left\langle f[\vec{X}_1]\right\rangle_X=\int d\vec{X}_1  P^{(M)}_{X}[\vec{X}_1 ]\,f[\vec{X}_1]\,,
\end{eqnarray}
where $f[\vec{X}_1]$ is an arbitrary function of $\vec{X}_1$. Using the notation (\ref{AverNotqtion}) the last term in (\ref{SMMutualinformation}) can be written as:
\begin{eqnarray}\label{loglambda}
\int d\vec{X}_1  P^{(M)}_{X}[\vec{X}_1 ]\,\log \Lambda_1[\vec{X}_1 ]=\langle \log
\Lambda_1 \rangle_{X}.
\end{eqnarray}
Since the PDF  (\ref{SMPXGaussian}) has the Gaussian form we have the following correlator
\begin{eqnarray}\label{Xpairing}
\left\langle X(\omega_k) \overline{X}(\omega_{k'}) \right\rangle_{X}=P
\delta_{k,\,k'}/\delta_{\omega},\qquad \omega_k=-W/2+ 2 \pi \,\delta_{\omega}
k,\qquad k,\, k'=0,1,\ldots,M-1.
\end{eqnarray}

From the representation (\ref{PathIntegral3}) we obtain the following expression for
$\Lambda_1[\vec{X}_1 ]$ in the frequency domain
\begin{eqnarray}\label{PathIntegral4}
&&\Lambda_1[\vec{X}_1 ]=\int\limits_{\phi (0,\omega)=0}^{\phi
(L,\omega)=0}\!\!\!\!\!\Big[{\cal D}\phi(z,\omega) \Big]_{M}\,
\exp\left\{-\frac{{\delta}_{\omega}}{Q}\!\!\sum^{M-1}_{k=0}\! \Delta
\sum^{N-1}_{n=1} {\cal L}_{eff}[\phi(z_n,\omega_{k})] \right\},
\end{eqnarray}
where the measure reads:
\begin{eqnarray} \label{measureintMomega}
\Big[{\cal D}\phi(z,\omega) \Big]_{M}=\lim_{\delta_{\omega} \rightarrow
0}\lim_{\Delta \rightarrow 0} \Big( \frac{\delta_{\omega}}{\Delta \pi
Q}\Big)^{M}\prod^{N-1}_{n=1} \prod^{M-1}_{k=0}\Big\{ \frac{\delta_{\omega}}{\Delta
\pi Q}\,\, dRe\phi(z_n,{\omega}_{k})\,dIm\phi(z_n, {\omega}_{k})\Big\}.
\end{eqnarray}

Now we present the Lagrangian ${\cal L}_{eff}$ as a sum:
\begin{eqnarray} \label{Lagrangianeff}
{\cal L}_{eff}={\cal L}^{(0)}_{eff}+{\cal L}^{(1)}_{eff}+{\cal L}^{(2)}_{eff}.
\end{eqnarray}
Here the first term is the leading order term in $\widetilde{\gamma}$ and it reads
\begin{eqnarray}\label{Leffzero}
&&{\cal L}^{(0)}_{eff}[\phi(z_n,\omega_{k})]=\left|(\partial_z -i\beta
\bar{\Omega}^2_{k}) \phi(z_n,\omega_{k})\right|^2,
\end{eqnarray}
where we have introduced the notation
\begin{eqnarray}\label{Omegabar}
&&\bar{\Omega}_{k}=2\sin[\pi k/ M]{M}/{T}=2 M \delta_{\omega} \sin[\pi k/
M]=W\sin[\pi k/ M]/\pi,
\end{eqnarray}
see Eq.~(\ref{SMOmega}). In the continuous limit $M \rightarrow \infty$ we will
assume that $\bar{\Omega}_{k}\approx 2\pi \delta_{\omega} k=\omega_{k}+W/2$. The
second term in the Lagrangian (\ref{Lagrangianeff}) reads
\begin{eqnarray}\label{Leffirst}
&&{\cal L}^{(1)}_{eff}[\phi(z_n,\omega_{k})]=2 \gamma Im \Big\{\left[ (\partial_z +i\beta
\bar{\Omega}^2_{k}) \bar{\phi}(z_n,\omega_{k}) \right]\,\delta^2_{\omega}
\sum^{M-1}_{k_1=0}\sum^{M-1}_{k_2=0}\sum^{M-1}_{k_3=0}
\Delta_{(M)}(k_1+k_2-k_3-k)\times \nonumber \\&& \left[ 2 \phi(z_n,\omega_{k_1})
\Phi(z_n,\omega_{k_2}) \bar{\Phi}(z_n,\omega_{k_3})+\bar{\phi}(z_n,\omega_{k_3})
\Phi(z_n,\omega_{k_1})\Phi(z_n,\omega_{k_2})\right]\Big\},
\end{eqnarray}
where $\Delta_{(M)}(k_1+k_2-k_3-k)$ is defined in Eq.~(\ref{SMdiscretedelta}). In
the following calculation we will use the function $\Phi(z_n,\omega_{k})$ in the
leading and next-to-leading order in $\gamma$:
$\Phi(z,\omega_{k})\approx\Phi^{(0)}(z,\omega_{k})+\Phi^{(1)}(z,\omega_{k})$, where
\begin{eqnarray}\label{Phi0}
&&\Phi^{(0)}(z,\omega_{k'}) = e^{i \beta \bar{\Omega}^2_{k'} z} X(\omega_{k'})\,,\qquad k'=0,1,\ldots,M-1,\\
\label{Phi1} &&\Phi^{(1)}(z,\omega_{k'}) = i \gamma e^{i \beta \bar{\Omega}^2_{k'}
z} \delta^2_{\omega}
\sum^{M-1}_{k'_1=0}\sum^{M-1}_{k'_2=0}\sum^{M-1}_{k'_3=0}X(\omega_{k'_1})X(\omega_{k'_2})
\bar{X}(\omega_{k'_3}) \Delta_{(M)}(k'_1+k'_2-k'_3-k') K(\mu,z),
\end{eqnarray}
where $\bar{\Omega}_{k'}$  is defined in Eq.~(\ref{Omegabar}),
$\mu=\mu(k',k'_1,k'_2,k'_3)=i \beta L
(\bar{\Omega}^2_{k'}+\bar{\Omega}^2_{k'_3}-\bar{\Omega}^2_{k'_1}-\bar{\Omega}^2_{k'_2})$,
and $K(\mu,z)=\left(1-e^{-\mu z/L}\right)/\mu$.

The third term in  the Lagrangian reads
\begin{eqnarray}\label{Leffsecond}
&&{\cal L}^{(2)}_{eff}[\phi(z_n,\omega_{k})]=\gamma^2 \delta^2_{\omega}
\sum^{M-1}_{k_1=0}\sum^{M-1}_{k_2=0}\sum^{M-1}_{k_3=0} \Delta_{(M)}(k_1+k_2-k_3-k)
\delta^2_{\omega} \sum^{M-1}_{k'_1=0}\sum^{M-1}_{k'_2=0}\sum^{M-1}_{k'_3=0}
\Delta_{(M)}(k'_1+k'_2-k'_3-k)\times \nonumber \\&& \left(2 \phi(z_n,\omega_{k_1})
\Phi(z_n,\omega_{k_2}) \bar{\Phi}(z_n,\omega_{k_3})+\bar{\phi}(z_n,\omega_{k_3})
\Phi(z_n,\omega_{k_1})\Phi(z_n,\omega_{k_2})\right)\times \nonumber \\&& \left[2
\bar{\phi}(z_n,\omega_{k'_1}) \bar{\Phi}(z_n,\omega_{k'_2})
{\Phi}(z_n,\omega_{k'_3})+{\phi}(z_n,\omega_{k'_3})
\bar{\Phi}(z_n,\omega_{k'_1})\bar{\Phi}(z_n,\omega_{k'_2})\right].
\end{eqnarray}
Let us stress that everywhere the derivatives with respect to $z$, i.e. $\partial_z
\phi(z_n,\omega_{k})$ encountered with other fields $\phi(z_n,\omega_{k})$ in the
same point $z_n=n \Delta$, are assumed as the difference derivatives in the
``causative'' manner: $f[\partial_z \phi(z_n,\omega_{k});\phi(z_n,\omega_{k})] =
f[\left( \phi(z_{n+1},\omega_{k})- \phi(z_n,\omega_{k})
\right)/\Delta;\phi(z_n,\omega_{k})]$, as provided by our approach
\cite{Terekhov:2014}. In what follows for brevity sake we will write a sum over $z$
($ \Delta \sum^{N-1}_{n=1} \ldots$) as an integral ($\int^L_0 dz \ldots$).

We present the perturbative expansion in $\widetilde{\gamma}$ of the normalization
factor $\Lambda_1$, see Eq.~(\ref{PathIntegral4}), in the form:
\begin{eqnarray}\label{lambdaexpansion}
\Lambda_1=\Lambda^{(0)}_1+\Lambda^{(1)}_1+\Lambda^{(2)}_1 +{\cal O}
(\widetilde{\gamma}^3),
\end{eqnarray}
where $\Lambda^{(m)}_1$ is of order of $\widetilde{\gamma}^m$.

Thus the last term in the expression ~(\ref{SMMutualinformation}) for the mutual
information has the following expansion in $\widetilde{\gamma}$:
\begin{eqnarray}\label{loglambdaexpansion}
\langle \log \Lambda_1 \rangle_{X}=  \left\langle \log \Lambda^{(0)}_1
\right\rangle_{X}+ \left\langle \frac{\Lambda^{(1)}_1}{\Lambda^{(0)}_1}
\right\rangle_{X} + \left( \left\langle \frac{\Lambda^{(2)}_1}{\Lambda^{(0)}_1}
\right\rangle_{X} - \frac{1}{2} \left\langle
\left(\frac{\Lambda^{(1)}_1}{\Lambda^{(0)}_1}\right)^2 \right\rangle_{X} \right)+
{\cal O} (\widetilde{\gamma}^3).
\end{eqnarray}

\emph{The order $\gamma^0$.} Retaining only the first term (\ref{Leffzero}) in the
exponent in Eq.~(\ref{PathIntegral4}) we arrive at:
\begin{eqnarray}\label{lambdazeroresult}
\Lambda^{(0)}_1=\int\limits_{\phi (0,\omega)=0}^{\phi
(L,\omega)=0}\!\!\!\!\!\Big[{\cal D}\phi(z,\omega) \Big]_{M}\,
\exp\left\{-\frac{{\delta}_{\omega}}{Q}\!\!\sum^{M-1}_{k=0}\! \int^L_0 dz
\left|(\partial_z -i\beta \bar{\Omega}^2_{k}) \phi(z,\omega_{k})\right|^2
\right\}=\left( \frac{\delta_{\omega}}{\pi Q L}\right)^{M}.
\end{eqnarray}
By taking into account that for the Gaussian distribution (\ref{SMPXGaussian}) the
input entropy has the form
\begin{eqnarray} \label{SMHXGaussian}
&&H[X]= M-M\log \left[ \delta_{\omega}/(\pi P)\right],
\end{eqnarray}
the expression for $\log \Lambda^{(0)}_1$ results in  the leading (Shannon's)
contribution $M \log {P/(QL)}$  to the mutual information.

For the path-integral (\ref{PathIntegral4}) we introduce the averaging $\langle
\ldots \rangle_{\phi}$ over fields $\phi(z,\omega)$  defined as
\begin{eqnarray}\label{averagingphi}
\langle( \ldots) \rangle_{\phi}=\frac{1}{\Lambda^{(0)}_1} \int\limits_{\phi
(0,\omega)=0}^{\phi (L,\omega)=0}\!\!\!\!\!\Big[{\cal D}\phi(z,\omega) \Big]_{M}\,
(\ldots) \exp\left\{-\frac{{\delta}_{\omega}}{Q}\!\!\sum^{M-1}_{k=0}\! \int^L_0 dz
\left|(\partial_z -i\beta \bar{\Omega}^2_{k}) \phi(z,\omega_{k})\right|^2 \right\}.
\end{eqnarray}
The paired correlator for this averaging can be calculated explicitly:
\begin{eqnarray}
&&\langle \phi(z,\omega_{k}) \overline{\phi}(z',\omega_{k'}) \rangle_{\phi}= -
\frac{Q}{\delta_{\omega}} \delta_{k,k'}G(z,z')\exp\left[i \beta \bar {\Omega}^2_{k}
(z-z')\right], \qquad  \langle \phi(z,\omega_{k}) {\phi}(z',\omega_{k'})
\rangle_{\phi}=0, \label{SMphipairingM}
\end{eqnarray}
where
\begin{eqnarray}\label{GreenF}
G(z,z')=z\dfrac{z'-L}{L}\, \theta(z'-z)+z'\dfrac{z-L}{L}\, \theta(z-z')
\end{eqnarray}
is the Green function of the  operator $\partial^2_{z}$ with the boundary conditions
$G(0,z')=G(L,z')=0$. In the notations (\ref{averagingphi}) we can
present the mutual information (\ref{SMMutualinformation}) as a sum of Shannon's
contribution and the term which is accountable for the impact of nonlinearity:
\begin{eqnarray}\label{SMMutualinformation1}
 I_{P[X]}=M \log\left[ P/(QL)\right]+ \left\langle \log   \left\langle \exp\left\{-\frac{S_{nl}
[\phi(z,\omega)]}{Q} \right\} \right\rangle_{\phi} \right\rangle_{X}, \quad
S_{nl}[\phi(z,\omega)]=\int^{L}_{0} dz \delta_{\omega} \sum^{M-1}_{k=0} \left( {\cal
L}^{(1)}_{eff}+{\cal L}^{(2)}_{eff} \right).
\end{eqnarray}

\emph{The order $\gamma^1$.} For the second term in the r.h.s. of
Eq.~(\ref{loglambdaexpansion}) we have
\begin{eqnarray}\label{lambda1result1}
\left\langle \frac{\Lambda^{(1)}_1}{\Lambda^{(0)}_1}  \right\rangle_{X}= - \left\langle
\frac{1}{Q} \int^{L}_{0} dz \delta_{\omega} \sum^{M-1}_{k=0}  \left\langle {\cal
L}^{(1)}_{eff}[\phi(z,\omega_k)]\Big|_{\Phi=\Phi^{(0)}} \right\rangle_{\phi}
\right\rangle_{X},
\end{eqnarray}
where ${\cal L}^{(1)}_{eff}$ is given by the expression (\ref{Leffirst}) with the
function $\Phi(z,\omega_k)$ being replaced with the zero order term
$\Phi^{(0)}(z,\omega_k)$, see Eq.~(\ref{Phi0}). Performing the averaging $\langle
\ldots \rangle_{\phi}$ by virtue of Eq.~(\ref{SMphipairingM}) we obtain that the
contribution (\ref{lambda1result1}) vanishes as the imaginary part of the real
value:
\begin{eqnarray}\label{lambda1result2}
\left\langle \frac{\Lambda^{(1)}_1}{\Lambda^{(0)}_1}  \right\rangle_{X}= 0.
\end{eqnarray}
It means that there are no corrections to the mutual information
(\ref{SMMutualinformation}) of order of $\widetilde{\gamma}$.

\emph{The order $\gamma^2$.} Let us consider the first term in Eq.~(\ref{loglambdaexpansion}) of order of $\gamma^2$.
There are three contributions to the quantity $\Lambda^{(2)}_1$:
\begin{eqnarray}\label{lambda2sum}
\Lambda^{(2)}_1=\Lambda^{(2.1)}_1+\Lambda^{(2.2)}_1+\Lambda^{(2.3)}_1.
\end{eqnarray}
The first contribution $\Lambda^{(2.1)}_1$ in (\ref{lambda2sum}) comes from the next-to-leading order expansion of the function $\Phi(z,\omega)$  in the expression (\ref{Leffirst}) for ${\cal L}^{(1)}_{eff}$: see Eq.~(\ref{Phi1}). For this contribution we can write the following representation
\begin{eqnarray}\label{lambda2-1}
&& \!\!\!  \frac{\Lambda^{(2.1)}_1}{\Lambda^{(0)}_1}=-\frac{4 \gamma}{ Q}
Im\,\Bigg\langle \int^{L}_{0} dz \,\delta_{\omega} \sum^{M-1}_{k=0}  (\partial_z +i\beta
\bar{\Omega}^2_{k}) \bar{\phi}(z,\omega_{k}) \,\delta^2_{\omega}
\sum^{M-1}_{k_1=0}\sum^{M-1}_{k_2=0}\sum^{M-1}_{k_3=0}
\Delta_{(M)}(k_1+k_2-k_3-k)\times \nonumber \\&&  \!\!\!\!\!\!\! \left[
\phi(z,\omega_{k_1})
\Big(\Phi^{(0)}(z,\omega_{k_2})\bar{\Phi}^{(1)}(z,\omega_{k_3})+
\Phi^{(1)}(z,\omega_{k_2})\bar{\Phi}^{(0)}(z,\omega_{k_3})\Big)
+\bar{\phi}(z,\omega_{k_3})
\Phi^{(0)}(z,\omega_{k_1})\Phi^{(1)}(z,\omega_{k_2})\right] \Bigg\rangle_{\phi},
\end{eqnarray}
where ${\Phi}^{(0)}$ is given by  Eq.~ (\ref{Phi0}),  and ${\Phi}^{(1)}$ is given by
 Eq.~ (\ref{Phi1}). After averaging $\langle\ldots \rangle_{\phi}$ with the help
of Eq.~(\ref{SMphipairingM}) we use Eqs.~(\ref{Phi0}) and  (\ref{Phi1}) to find
$\Big\langle {\Lambda^{(2.1)}_1}/{\Lambda^{(0)}_1} \Big\rangle_{X}$. From the Wick
theorem \cite{IZ:1980,Peskin:1995} with the correlator (\ref{Xpairing}) applied to
the averaging over $X$ we find that all pairings cancel with each other:
\begin{eqnarray}\label{lambda2-1result}
&&\left\langle \frac{\Lambda^{(2.1)}_1}{\Lambda^{(0)}_1} \right\rangle_{X}=0.
\end{eqnarray}
The second contribution $\Lambda^{(2.2)}_1$ in (\ref{lambda2sum}) comes from  ${\cal
L}^{(2)}_{eff}$, see Eq.~(\ref{Leffsecond}), where $\Phi(z,\omega)$ is considered in
the leading order in $\gamma$, see Eq.~(\ref{Phi0}). For this contribution we can write the
following representation
\begin{eqnarray}\label{lambda2-2}
&& \!\!\!  \frac{\Lambda^{(2.2)}_1}{\Lambda^{(0)}_1}=- \frac{1}{Q} \int^{L}_{0} dz
\delta_{\omega} \sum^{M-1}_{k=0}  \left\langle {\cal
L}^{(2)}_{eff}[\phi(z,\omega_k)]\Big|_{\Phi=\Phi^{(0)}}\right\rangle_{\phi}.
\end{eqnarray}
The averaging $\langle\ldots \rangle_{\phi}$ is straightforward.  To find the
quantity  $\Big\langle {\Lambda^{(2.2)}_1}/{\Lambda^{(0)}_1} \Big\rangle_{X}$ we use
the Wick theorem \cite{IZ:1980,Peskin:1995} with the correlator
Eq.~(\ref{Xpairing}). After Wick pairing we verify that the dependence on $\beta$
vanishes for this quantity. Finally, we have
\begin{eqnarray}\label{lambda2-2result}
\left \langle \frac{\Lambda^{(2.2)}_1}{\Lambda^{(0)}_1}
\right\rangle_{X}=-\frac{5}{3}M \widetilde{\gamma}^2.
\end{eqnarray}
Here we have used the value of the integral $\int^L_0 G(z,z)dz= -{L^2}/{6} $ with
the Green function  (\ref{GreenF}). When obtaining (\ref{lambda2-2result}) we also
used that $\sum^{M-1}_{k=0} \Delta_{(M)}(k_1+k_2-k_3-k)=1$, where $0 \leq k_i \leq
M-1$, see Eq.~(\ref{SMdiscretedelta}).

The third contribution $\Lambda^{(2.3)}_1$ in (\ref{lambda2sum}) comes from the
expansion of the exponent in Eq.~(\ref{SMMutualinformation1}) of order of
$\left[{\cal L}^{(1)}_{eff}\right]^2$:
\begin{eqnarray}\label{lambda2-3}
&& \!\!\! \frac{\Lambda^{(2.3)}_1}{\Lambda^{(0)}_1}= \frac{1}{2 Q^2} \int^{L}_{0}
dz_1 \int^{L}_{0} dz_2\, \delta^2_{\omega} \sum^{M-1}_{k=0} \sum^{M-1}_{k'=0}
\left \langle {\cal
 L}^{(1)}_{eff}[\phi(z_1,\omega_k)]{\cal L}^{(1)}_{eff}[\phi(z_2,\omega_k')]
\Big|_{\Phi=\Phi^{(0)}}\right \rangle_{\phi},
\end{eqnarray}
where ${\cal L}^{(1)}_{eff}$ appears in Eq.~(\ref{lambda2-3}) through the
representation (\ref{Leffirst}) with $\Phi(z,\omega)$ being replaced with the
leading order term $\Phi^{(0)}(z,\omega_k)$, see Eq.~(\ref{Phi0}). The calculation
of the contribution $ \Big\langle {\Lambda^{(2.3)}_1}/{\Lambda^{(0)}_1}
\Big\rangle_{X}$ to the r.h.s. of Eq.~(\ref{loglambdaexpansion}) is the most
cumbersome but straightforward. Here we present the result of the calculation:
\begin{eqnarray}\label{lambda2-3result}
&& \left \langle \frac{\Lambda^{(2.3)}_1}{\Lambda^{(0)}_1} \right \rangle_{X}= M
\widetilde{\gamma}^2\left[\frac{5}{3}-\frac{1}{3}g(\widetilde{\beta}) \right],
\end{eqnarray}
where the function $g(\widetilde{\beta})$ ($\widetilde{\beta}=\beta L W^2$ is
dimensionless dispersion parameter) can be presented in our discretization scheme in
the form of a triple sum:
\begin{eqnarray}\label{SMg-function1}
&&g(\widetilde{\beta})=  \frac{1}{M^3} \sum^{M-1}_{k_1=0} \sum^{M-1}_{k_2=0}
\sum^{M-1}_{k_3=0} F\left(\frac{\widetilde{\beta}}{2}\Big[
\bar{\Omega}^2_{k_1}+\bar{\Omega}^2_{k_2}-\bar{\Omega}^2_{k_3}-\bar{\Omega}^2_{k_1+k_2-k_3}
\Big]\right).
\end{eqnarray}
The function $F(\mu)$ in Eq.~(\ref{SMg-function1}) is the result of integration of
the derivatives of the dimensionless Green function, $G_0(\zeta_1,\zeta_2)=\zeta_1
(\zeta_2-1)\theta(\zeta_2-\zeta_1)+\zeta_2 (\zeta_1-1)\theta(\zeta_1-\zeta_2)$, see
Eq.~(\ref{GreenF}),
\begin{eqnarray}\label{F-function0}
&&F(\mu)=-12 \int^{1}_{0}d\zeta_1 \int^{1}_{0}d\zeta_2 \frac{\partial
G_0(\zeta_1,\zeta_2)}{\partial \zeta_1}  \frac{\partial
G_0(\zeta_1,\zeta_2)}{\partial \zeta_2} e^{-2 i \mu (\zeta_1 -\zeta_2
)}=3\frac{\mu^2-\sin^2(\mu)}{\mu^4}=4!\sum^{\infty}_{s=0} \frac{(-1)^s
(2\mu)^{2s}}{(2s+4)!},
\end{eqnarray}
and for convenience it is normalized as $F(0)=1$.

The last term of order of $\gamma^2$ in Eq.~(\ref{loglambdaexpansion})  is zero for
the same reasons as in Eq.~(\ref{lambda1result2}).
\begin{eqnarray}\label{lambda2result3}
\left\langle \left(\frac{\Lambda^{(1)}_1}{\Lambda^{(0)}_1}\right)^2 \right \rangle_{X}= 0.
\end{eqnarray}

Now we call together all terms in Eq.~(\ref{loglambdaexpansion}): Eqs.
(\ref{lambdazeroresult}), (\ref{lambda1result2}), (\ref{lambda2-1result}),
(\ref{lambda2-2result}), (\ref{lambda2-3result}), and (\ref{lambda2result3}).
Finally, we obtain the following expression for the mutual information in the
leading order in $1/\mathrm{SNR}$ for the Gaussian PDF (\ref{SMPXGaussian}):
\begin{eqnarray}
I_{P_G[X]}=M \left\{ \log \mathrm{SNR}-  \frac{\widetilde{\gamma}^2}{3}
g(\widetilde{\beta}) \right\}+{\cal O}(\widetilde{\gamma}^{\,4}),
\label{SMMutInfRes}
\end{eqnarray}
where $M$ is the number of complex meaning channels. This number is implied to be
large $M \gg 1$. In the continuous limit of sufficiently large $M$ we present
$g(\widetilde{\beta})$ through the integral:
\begin{eqnarray}\label{g-function3}
&&g(\widetilde{\beta})=\int^{1}_{0}dx_1 \int^{1}_{0}dx_2 \int^{1}_{0}dx_3
F\left( \frac{\widetilde{\beta}}{4}(x_1-x_3)(x_2-x_3)\right),
\end{eqnarray}
where we have used the expansion of $\bar{\Omega}_{k}$, see Eq.~(\ref{Omegabar}),
when the argument of the sinus is close to zero or to $\pi$, and we have used the
identity $x^2_1+x^2_2-x^2_3-(x_1+x_2-x_3)^2=-2(x_1-x_3)(x_2-x_3)$. To calculate
analytically the integral (\ref{g-function3}) one can perform the series expansion
(\ref{F-function0}) for $F(\mu)$ and the term by term integration for the
polynomials $(x_1-x_3)^{2n}(x_2-x_3)^{2n}$:
\begin{eqnarray}\label{g-function4}
&&g(\widetilde{\beta})=4!\sum^{\infty}_{n=0} a_n \frac{(-1)^n
(\widetilde{\beta}/2)^{2n}}{(2n+4)!},\qquad   \\&& a_n=\int^{1}_{0}dx_1
\int^{1}_{0}dx_2 \int^{1}_{0}dx_3
(x_1-x_3)^{2n}(x_2-x_3)^{2n}=\frac{2}{(1+2n)^2}\left(\frac{1}{3+4n} +
\frac{\Gamma(2+2n)^2}{\Gamma(4+4n)}\right) \nonumber.
\end{eqnarray}
The series (\ref{g-function4}) can be calculated in the term of the generalized
hypergeometric functions.


For $\widetilde{\beta}=0$ we have $g(0)=1$ and we arrive at the
result \cite{TTKhR:2015} for the nondispersive channel in expansion in
$\widetilde{\gamma}$:
\begin{eqnarray}
I_{P[X]}=M \log \mathrm{SNR}- M \frac{\widetilde{\gamma}^2}{3}+{\cal
O}(\widetilde{\gamma}^{\,4}). \label{MutInfResbetazero}
\end{eqnarray}

For large $\widetilde{\beta}$ we can consider the asymptotics of the function
$g(\widetilde{\beta})$ obtained from the integral representation
(\ref{g-function3}):
\begin{eqnarray}\label{SMg-functionlarge}
&&g(\widetilde{\beta})= \frac{16 \pi}{\widetilde{\beta}}\left( \log
\widetilde{\beta} -\log 2+\gamma_{E} -\frac{23}{6}\right)+{\cal O}\left(
\frac{1}{\widetilde{\beta}^{3/2}}\right), \qquad \widetilde{\beta} \rightarrow
\infty.
\end{eqnarray}
